\documentclass[12pt]{article}
\usepackage{graphicx}
\usepackage{amssymb}
\usepackage{amsthm}
\usepackage{amsmath}
\usepackage[usenames]{color}

\makeatletter
\@addtoreset{equation}{section}

\makeatletter
\renewcommand\section{\@startsection {section}{1}{\z@}%
                                   {-3.5ex \@plus -1ex \@minus -.2ex}
                                   {2.3ex \@plus.2ex}%
                                   {\normalfont\large\bfseries}}
\renewcommand\subsection{\@startsection{subsection}{2}{\z@}%
                                     {-3.25ex\@plus -1ex \@minus -.2ex}%
                                     {1.5ex \@plus .2ex}%
                                     {\normalfont\bfseries}}

\newcommand{\be}{\begin{equation}}
\newcommand{\ee}{\end{equation}}
\newcommand{\beq}{\begin{eqnarray}}
\newcommand{\eeq}{\end{eqnarray}}

\def\[{\left [}
\def\]{\right ]}
\def\({\left (}
\def\){\right )}

\def\benum{\begin{enumerate}}
\def\eenum{\end{enumerate}}
\def\bitem{\begin{itemize}}
\def\eitem{\end{itemize}}

\def\btab{\begin{table}[h] \begin{center} \begin{tabular}{l lp{3in}}}
\def\etab{\end{tabular} \end{center} \end{table}}
\def\btabm{\begin{center} \begin{tabular}}
\def\etabm{\end{tabular} \end{center}}

\def\f#1#2{{\frac{#1}{#2}}}

\def\p{\partial}

\def\m{\cdot}

\def\half{{1\over 2}}
 
 \def\m{{\mu}}
 
 \def\n{{\nu}}
 \def\ep{{\epsilon}}

 \def\frac#1#2{{#1\over #2}}

 \def\D{{\Delta}}
 \def\g{{\gamma}}
 \def\s{\sqrt}

 \def\p{\partial}

 \def\al{\alpha'}
 \def\de{\partial}

 \def\f {\frac}
 \def\ti{\tilde}
 \def\ap{\alpha}

 \def\ddd{\cdot\cdot\cdot}
 \def\no{\nonumber \\}
 \def\la{\langle}
 \def\lb{\rangle}
 \def\ep{\epsilon}

 \def\vp{\varphi}
\def\ba{\begin{eqnarray}}
\def\ea{\end{eqnarray}}

\def\half{{1\over 2}}
 
 \def\m{{\mu}}
 
 \def\n{{\nu}}
 \def\ep{{\epsilon}}

 \def\frac#1#2{{#1\over #2}}

 \def\D{{\Delta}}
 \def\g{{\gamma}}
 \def\s{\sqrt}

 \def\p{\partial}

\begin{document}

\begin{titlepage}
\thispagestyle{empty}


KUNS-2105
\begin{flushright}
\end{flushright}


\begin{center}
\noindent{\large {\textbf{Near Extremal Black Hole Entropy }\\
\textbf{as Entanglement Entropy via AdS$_2$/CFT$_1$}}}

 \vspace{2cm}
\noindent{Tatsuo
Azeyanagi\footnote{e-mail:aze@gauge.scphys.kyoto-u.ac.jp},
 Tatsuma Nishioka\footnote{e-mail:nishioka@gauge.scphys.kyoto-u.ac.jp} and Tadashi
Takayanagi\footnote{e-mail:takayana@gauge.scphys.kyoto-u.ac.jp}}\\
\vspace{1cm}

 {\it  Department of Physics, Kyoto University, Kyoto 606-8502, Japan }

\vskip 2em
\end{center}

\begin{abstract}

We point out that the entropy of (near) extremal black holes can be
interpreted as the entanglement entropy of dual conformal quantum
mechanics via AdS$_2/$CFT$_1$. As an explicit example, we study near
extremal BTZ black holes and derive this claim from AdS$_3/$CFT$_2$.
We also analytically compute the entanglement entropy in the two
dimensional CFT of a free Dirac fermion compactified on a circle at
finite temperature. From this result, we clarify the relation
between the thermal entropy and entanglement entropy, which is
essential for the entanglement interpretation of black hole entropy.

\end{abstract}

\end{titlepage}

\newpage

\section{Introduction}

\hspace{5mm} The AdS/CFT \cite{Maldacena} has been studied for
almost ten years and many interesting aspects of quantum gravity
have been revealed. Even though the examples of
AdS$_{d+1}/$CFT$_{d}$ with $d\geq 2$ have been explored in detail,
the lowest dimensional case $d=1$ has not been understood well even
now. The AdS$_2$ geometry appears as the near horizon limit of four
or five dimensional extremal black holes (or black rings) \cite{KLR,
AGJST,AY}. Thus the microscopic explanation of Bekenstein-Hawking
entropy of the extremal black holes \cite{SV,CaMa} is expected to be
directly related to the AdS$_2/$CFT$_1$ correspondence
\cite{Maldacena,St} (see \cite{BMSV} for a review).

The pure AdS spacetime AdS$_{d+1}$ with $d\geq 2$ has no entropy as
is also clear from its dual CFT$_{d}$ at vanishing temperature. To
obtain non-zero entropy, we need to consider an AdS black hole as
the dual geometry. On the other hand,
 we expect non-zero entropy even for the pure AdS$_2$ since it is the
near horizon limit of higher dimensional (near) extremal black
holes. We also notice another special property of AdS$_2$ that the
AdS$_{d+1}$ in the global coordinate has two (time-like) boundaries
only when $d=1$. The latter property has been a major problem when
we would like to understand what the AdS$_2/$CFT$_1$ is because
usually the CFT lives on the boundary of AdS space. So far this
issue has been neglected and the AdS$_2$ space is considered to be
dual to a single CFT with a large degeneracy. Though it is also
natural to assume that there are two CFTs, taking into account the
presence of two boundaries on AdS$_2$, there have been no arguments
in this direction as far as the authors know.

In this paper, we would like to report a progress in this direction
owing to the recently found method of holographically computing
entanglement entropy \cite{RT,HRT}. We point out that the above two
exceptional properties of AdS$_2$ are closely related with each
other. We present an important evidence that there exist two systems
of conformal quantum mechanics (CQM) on the boundaries of the
AdS$_2$ and that they are entangled with each other as is speculated
from the non-vanishing correlation functions between them computed
holographically. Indeed we will be able to show that the black hole
entropy is exactly the same as the entanglement entropy of CQM if we
assume the AdS$_2/$CFT$_1$ correspondence with this interpretation.
This relation is true even if we take any higher derivative
corrections into account. We can say that this progress is highly
remarkable if we remember that the AdS$_2/$CFT$_1$ has been poorly
understood and been still mysterious until now.

Even though our argument can be regarded as a generalization of the
interpretation of AdS black holes in \cite{ME} via AdS/CFT, it is
slightly different from it in the following point. For the (3D or
higher dimensional) AdS black holes, its CFT dual is well
established and it is possible to explicitly construct a dual
entangled CFT state, from which we can compute its entanglement
entropy directly as in \cite{ME}. On the other hand, in the AdS$_2$
case, we can perform a computation of entanglement entropy in the
dual CFT only by using the recent holographic method\footnote{This
holographic method has also been applied to the analysis of the
confining gauge theories \cite{KKM,NT}.} \cite{RT,HRT} as the
formulation of the dual CFT is not clear at present.

The relation between the black hole entropy and entanglement entropy
has been discussed for a long time and historically this was the
first motivation that makes us consider the entanglement entropy in
quantum field theory \cite{BK}. Later, it turned out that quantum
corrections to Bekenstein-Hawking formula can be explained as the
entanglement entropy \cite{SuUg,PrSt}. In particular, when the
entire gravity action is induced, the black hole entropy itself can
be regarded as the entanglement entropy \cite{PrSt,Ja}. In these
arguments, the black hole entropy is related to the entanglement
entropy {\it in the quantum field theory in the same spacetime}. The
corresponding interpretation from the viewpoint of AdS/CFT has been
given in \cite{HMS,Em} (see also \cite{Eh,So,Cad}). On the other
hand, in our case, the black hole entropy is interpreted as the
entanglement entropy {\it in CFT (or CQM) which lives on the
boundary of the spacetime} .

Now, we usually identify the black hole entropy with the thermal
entropy based on AdS/CFT. Thus in order to claim the equality
between the black hole entropy and the entanglement entropy in
general setups, we need to establish the relation between the
thermal entropy and entanglement entropy. To see that it indeed
agrees with what we expect from the holographic viewpoint, we
compute the entanglement entropy of a 2D free Dirac fermion at
finite temperature with the spatial direction compactified as an
explicit example. We finally obtain an analytical expression and are
able to check this relation. This is the first analytic result on
entanglement entropy with both the finite temperature and finite
size effect taken into account\footnote{Since this result may also
be interesting for those who are interested in other subjects, we
arranged such that the section 5 is readable for anyone who is
familiar with 2D CFT.}. Also remarkably, in our setup the
entanglement entropy depends on the detail of the 2D CFT, while the
entanglement entropy at zero temperature or in the infinite system
only depends on the central charge of CFT \cite{HLW,Cardy}.

This paper is organized as follows: In section 2 we explain the
holographic computation of entanglement entropy via the AdS/CFT
duality. We also presents a new evidence of this relation in the BTZ
black holes. In section 3, we give a general argument to show the
equivalence between the black hole entropy and the entanglement
entropy via AdS$_2/$CFT$_1$. In section 4, we investigate near
extremal BTZ black holes in order to derive our claim from
AdS$_3/$CFT$_2$. In section 5, we analytically compute the
entanglement entropy of a 2D free Dirac fermion at finite
temperature with the spacial direction compactified. In section 6,
we draw a conclusion and discuss future problems.

\section{Holographic Entanglement Entropy and BTZ Black Holes}

\hspace{5mm} The main purpose of this paper is to understand the
AdS$_2$/CFT$_1$ better by uncovering the relation between the black
hole entropy and the entanglement entropy in CFT$_1$. However, it is
quite useful to learn a general holographic prescription of
computing entanglement entropy from the AdS/CFT correspondence. This
is because the dual CFT in AdS$_2$/CFT$_1$ is not understood well
and we need to employ a holographic computation of the entanglement
entropy\footnote{Again please distinguish this entanglement entropy
in CFT$_1$ from the entanglement entropy in AdS$_2$.} for CFT$_1$.
We will apply this general method to the AdS$_2$/CFT$_1$ setup in
the next section. Also in a particular case of AdS$_2$ background in
string theory can be embedded into a rotating BTZ black hole, which
is asymptotically AdS$_3$ as we will see.

Motivated by this, we will explain the general holographic
computation of the entanglement entropy \cite{RT} in this section.
Especially we study the example of BTZ black holes and its CFT$_2$
dual based on the AdS$_3$/CFT$_2$ and present a new result. This
gives a further evidence that the general prescription in \cite{RT}
correctly reproduces the black hole entropy as the entanglement
entropy. Also the entropy of BTZ black hole is closely related to
the entropy of extremal black holes which is the main topic of this
paper as we will see later.

\subsection{Holographic Entanglement Entropy}\label{ss:HolIntgg}

\hspace{5mm} Consider a CFT and divide the space manifold of the CFT
into two parts $A$ and $B$. This factorizes the total Hilbert space
into a direct product of two Hilbert spaces $H_A\otimes H_B$. The
entanglement entropy is defined by the von-Neumann entropy
$S_A=-\mbox{Tr}\rho_A\log \rho_A$ for the reduced density matrix
$\rho_A$. The reduced density matrix $\rho_A$ is defined by tracing
out the density matrix over $H_B$ i.e. $\rho_A=\mbox{Tr}_{B}\rho$.
We have an infinitely many such quantities for various choices of
$A$.

Now we would like to compute the entanglement entropy from the
AdS/CFT correspondence. We assume a setup where a AdS$_{d+2}$ space
with the Newton constant $G^{(d+2)}_N$ is dual to a CFT$_{d+1}$. The
CFT lives on the boundary of AdS. Then the general holographic
prescription in \cite{RT} computes the entanglement entropy as the
area of the minimal surface at a constant time \be S_A =
\f{\mbox{Area}(\g_A)}{4G^{(d+2)}_N}, \label{ARE} \ee where
$\gamma_A$ is the (unique) minimal surface in AdS$_{d+2}$ whose
boundary coincides with the boundary of the region A. A simple proof
of this claim has been given in \cite{Fu}. Notice that this formula
assumes the supergravity approximation of the full string theory.

\subsection{Application to BTZ Black holes}\label{ss:HolInt}

\hspace{5mm} As a particular example, which is also relevant to the
discussions in the next section, let us consider the BTZ black holes
\cite{BTZ}, whose metric is given as follows \be
ds^2=-\f{(r^2-r_-^2)(r^2-r_+^2)}{R^2r^2}dt^2
+\f{R^2r^2}{(r^2-r_-^2)(r^2-r_+^2)}dr^2+r^2
\left(d\phi+\f{r_+r_-}{Rr^2}dt\right)^2.\label{metnon} \ee The
boundary of BTZ black hole at a fixed time is a circle because
$\phi$ has the periodicity $\phi\sim\phi+2\pi$. The entanglement
entropy is defined by dividing this circle into two parts $A$ and
$B$. We specify the size of $A$ by the angle $\Delta \phi=2\pi L$,
while the size of $B$ becomes $\Delta \phi=2\pi(1-L)$.

If we apply the holographic formula (\ref{ARE}) to BTZ black holes,
$\mbox{Area}(\g_A)$ is equal to the geodesic length between the two
endpoints of $A$ inside the bulk space. This holographic computation
leads to the following prediction \cite{RT} \be S_A=\f{c}{3}\log
\left[\f{\beta}{\pi a}\sinh\left(\f{\pi
L}{\beta}\right)\right],\label{entdb} \ee where $c$ is the central
charge of the dual CFT$_2$ and $\beta$ is the inverse temperature of
the black hole. This agrees with the result in \cite{Cardy}, which
computes the entanglement entropy in any 2D finite temperature CFT
when the size $L$ is small. However, when $L$ is large, the formula
(\ref{entdb}) is no longer correct as will be clear from the
holographic consideration discussed just below.

\begin{figure}[htbp]
\begin{tabular}{ccc}
\begin{minipage}{5.7cm}
  \begin{center}
      \includegraphics[scale=0.35,clip]{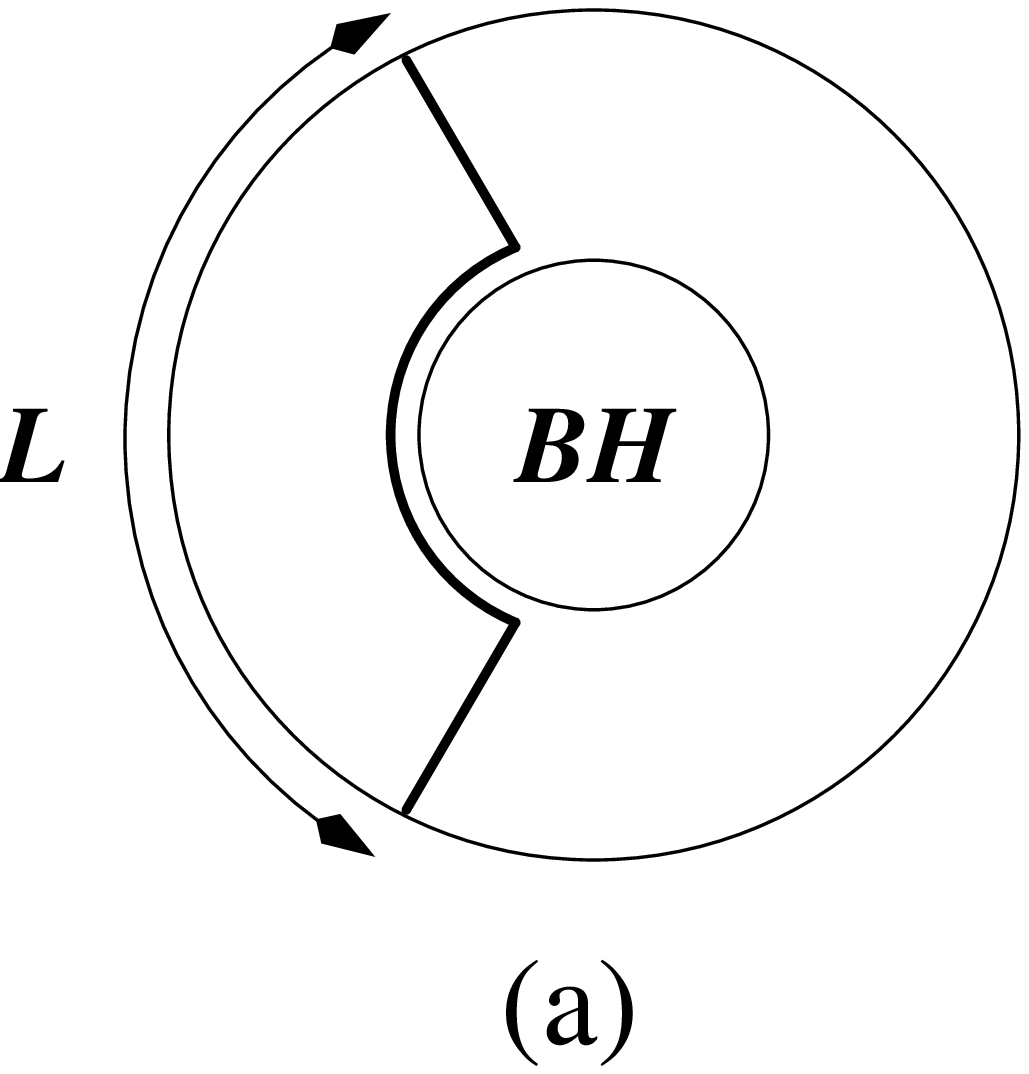}
  \end{center}
\end{minipage}
&
\begin{minipage}{5.5cm}
  \begin{center}
    \includegraphics[scale=0.35,clip]{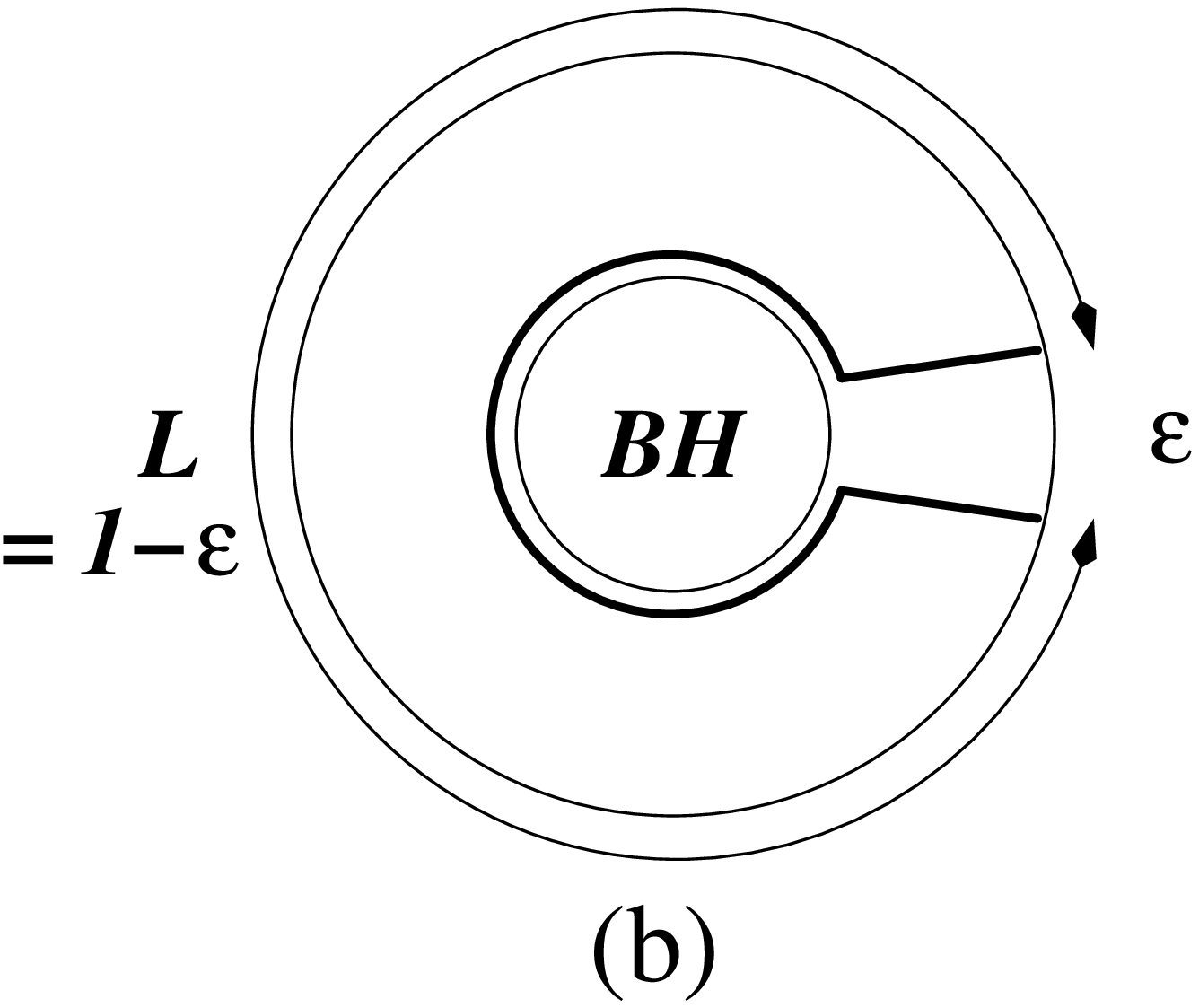}
  \end{center}
\end{minipage}
&
\begin{minipage}{5.5cm}
  \begin{center}
    \includegraphics[scale=0.35,clip]{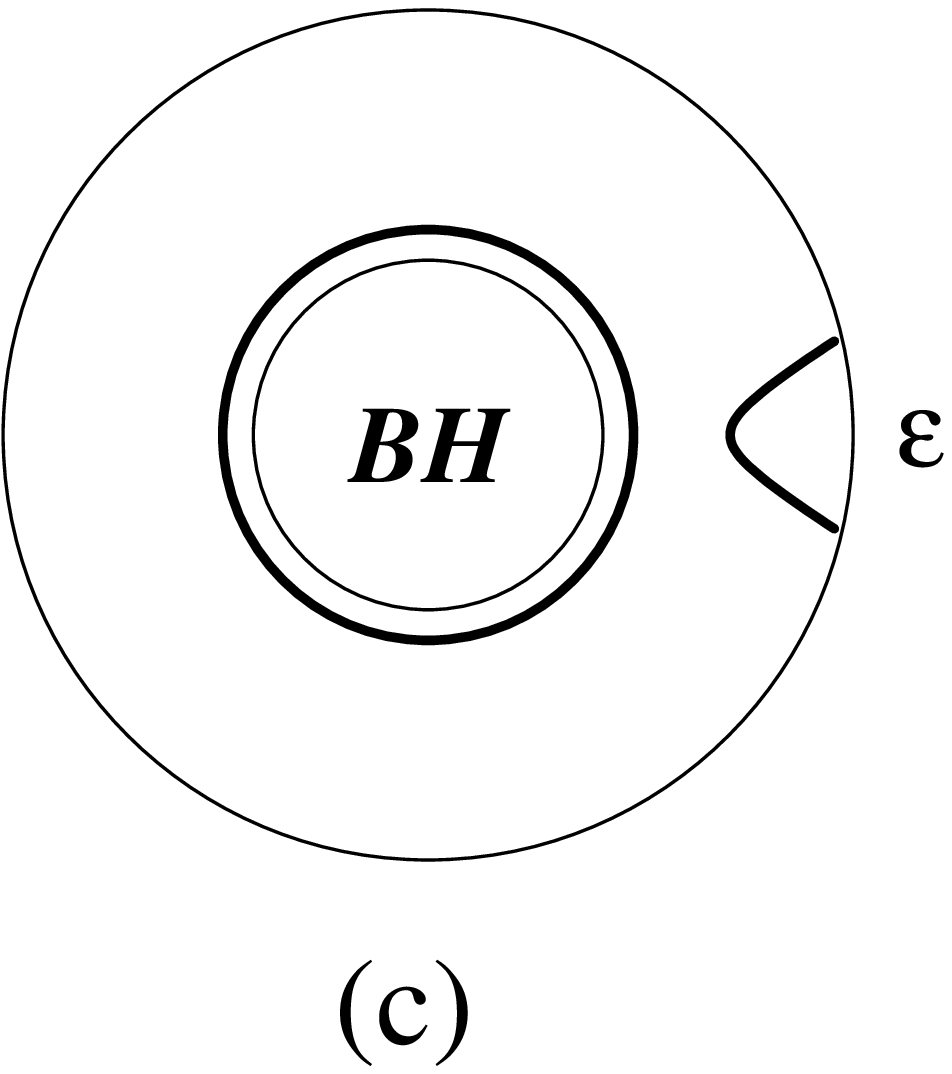}
  \end{center}
\end{minipage}
\end{tabular}
\caption{Holographic picture of the entanglement entropy. (a) The
length of the geodesic $\g_A$ whose boundary coincides with $\p A$
gives the holographic entanglement entropy of the region $A$. (b)
The region $A$ covers almost all the boundary as the length of the
region $A$ gets large. (c) The disconnected curves with the same
boundary as the one of (b), gives another candidate of $\g_A$. This
consists of a part of black hole horizon and the geodesic extending
to the boundary. The former has a finite length, while the latter is
infinitely long $\sim \f{c}{3}\log (\ep/a)$.} \label{fig:HolInt}
\end{figure}

At high temperature, the geodesic winds around the black hole
horizon as $L$
 becomes large (Figure \ref{fig:HolInt}(a)).
When the region $A$ covers most of the boundary ($L=1-\epsilon$ with
$\ep<<1$), the disconnected surface (Figure \ref{fig:HolInt}(c))
gives smaller area than\footnote{Remember that when the temperature
is non-vanishing but is not high enough, the AdS/CFT claims that the
dual gravity description is given by the path-integral over
infinitely many geometries as in \cite{MS,DMMV}. Thus our results
such as (\ref{toth}) and (\ref{LOW}), which are correct for any
values of $\beta$, should include such a sum over geometries. } the
connected surface (Figure \ref{fig:HolInt}(b)). Thus the
disconnected surface consists of the total black hole horizon and
the geodesic extending to the boundary. Taking the $\ep\to 0$ limit,
this leads to \be S_A(L=1-\ep) = S_{BH} + S_A(L=\ep),
\label{relatione} \ee where $S_{BH}$ is the black hole entropy. This
relation (\ref{relatione}) offers an important way to extract the
black hole entropy from the entanglement entropy of CFT$_2$.

Therefore it is very important to confirm (\ref{relatione}) from the
CFT side without assuming AdS/CFT. Indeed in section 5.3 we will
show this is indeed true for a particular CFT. There, we consider
the example\footnote{In this subsection, we have proceeded by
pretending that the free Dirac fermion system has its AdS dual. We
believe this assumption is not crucial because the property
(\ref{dif}) should be true for any 2D CFT. It is well-known that the
IIB string on AdS$_3\times S^3\times M$ ($M=K3$ or $T^4$) is dual to
the 2D $(4,4)$ SCFT defined by the symmetric orbifolds $Sym(M)^N$.
Thus it will be an interesting future problem to extend our
calculations of entanglement entropy to the ones in symmetric
orbifolds $Sym(M)^N$ and see that the result can explicitly be
interpreted as the sum over geometries.} of free fermion CFT since
it turns out to be possible to compute the entanglement entropy
analytically and show this relation as in (5.22).

In this way we have been able to understand well the BTZ black hole
entropy from the viewpoint of entanglement entropy. This gives a
further evidence of AdS$_3$/CFT$_2$. In the next sections, we would
like to proceed to another important class of black holes i.e. the
ones whose near horizon geometry includes the AdS$_2$.

\section{Black Hole Entropy as Entanglement Entropy and AdS$_2$/CFT$_1$}

\subsection{AdS$_2$ from the Near Horizon Limit of Extremal Black Hole}

\hspace{5mm} The metric of a 4D charged black hole looks like \be
ds^2=-\frac{(r-a_+)(r-a_-)}{r^2}dt^2+\frac{r^2}{(r-a_+)(r-a_-)}dr^2+r^2
d\Omega_2^2, \ee where we assumed $a_+\geq a_-$. The
Bekenstein-Hawking entropy is given by \be
S_{BH}=\frac{A}{4G^{(4)}_N}=\frac{\pi a_+^2}{G^{(4)}_{N}}. \ee The
extremal black hole corresponds to the special choice of the
parameter $a_+=a_-$. In this case, if we define $u=r-a_+$, the near
horizon metric becomes \be ds^2=-\f{u^2}{a_+^2}
dt^2+a_+^2\f{du^2}{u^2}+a_+^2 d\Omega_2^2, \ee i.e. AdS$_2$  in the
Poincare coordinate times $S^2$.

More generally, it is possible to obtain AdS$_2\times S^2$ when the
black hole is near extremal $\f{a_+-a_-}{a_+}\ll 1$ \cite{SS}. In
this case, the dual ground state in AdS$_2$ is heated up into a
thermal state so that its temperature is proportional to $a_+-a_-$.
As we will see in the last part of the next section, the extremal
black hole $a_+=a_-$ behaves differently from the near extremal one
especially in the global structure of the spacetime. Below we mostly
consider the extremal limit of the near extremal black hole instead
of the extremal one itself.

As is well known, the AdS$_2$ in the global coordinate \be
ds^2=a_+^2\f{-d\tau^2+d\sigma^2}{\cos^2\sigma}, \label{globaladst}
\ee has a significant difference from the higher dimensional AdS
spaces in that it has two time-like boundaries at $\sigma=\pm
\f{\pi}{2}$. Thus it is natural to expect that the theory is dual to
two copies of conformal quantum mechanics CQM$1$ and CQM$2$ living
on the two boundaries via AdS$_2$/CFT$_1$. In the next section, by
considering 5D (near) extremal black holes, we will give an explicit
example of AdS$_2$/CFT$_1$ duality, which supports this
interpretation.

In the case of 4D extremal black holes, the systematic construction
of dual CQM has not been established. There are some specific
examples whose dual quantum mechanics is understood
\cite{SQ,GSY,DM}. Instead of the detailed review of each examples,
we would like to briefly give a sketchy explanation since the detail
is not necessary for our purpose. Consider the setup of type IIA
string compactified on a Calabi-Yau 3-fold with D$0$-branes and
D$4$-branes. We specify the number of D$0$-branes and D$4$-branes
wrapped on the 4-cycle $\ap_A$ by $q_0$ and $p^A$. This
configuration leads to a macroscopic BPS black hole with the entropy
$S=2\pi\s{q_0D}$ in a large charge limit, where
$D=\f{1}{6}C^{ABC}p_Ap_Bp_c$ in terms of the intersection number
$C^{ABC}$ \cite{MSW}. In the near horizon limit, the geometry
AdS$_2\times S^2$ is realized. In this setup, the dual quantum
mechanics is described by a supersymmetric sigma model whose target
space is the symmetric product $Sym(P^{q_0})$ of a certain manifold
$P$ \cite{SQ}. This manifold $P$ represents the effective geometry
of D4-brane world-volume probed by a D$0$-brane. The number of
ground states $d(q_0)$ of this model is equal to the number of
cohomology of the symmetric product $Sym(P^{q_0})$. We can apply the
orbifold formula as usual to count $d(q_0)$ \cite{V,SV}. This turns
out to be equivalent to the counting of left-moving states of a
 two dimensional CFT at level $q_0$ with the central charge $c_L=6D$
\cite{SQ,MSW}. This reproduces $S=\log d(q_0)=2\pi\s{q_0D}$. In this
setup, we can regard the pair CQM1 and CQM2 as the two copies of the
symmetric product quantum mechanics.

\subsection{Holographic Computation of Entanglement Entropy}

\hspace{5mm} Since there are two CQMs, it is natural to ask if there
are any correlations between them. We can compute from the standard
bulk-boundary relation \cite{GKPW} the two point function between
${\cal{O}}_1$ in CFT$_1$ and ${\cal{O}}_2$ in CFT$_2$ as follows (we
assume the global AdS$_2$ (\ref{globaladst})) \ba \la
{\cal{O}}_1(\tau_1) {\cal{O}}_1(\tau_2)\lb
=\f{1}{\left[\sin\left(\f{\tau_1-\tau_2}{2}\right)\right]^{2h}},\label{onet}\\
\la {\cal{O}}_1(\tau_1) {\cal{O}}_2(\tau_2)\lb
=\f{1}{\left[\cos\left(\f{\tau_1-\tau_2}{2}\right)\right]^{2h}},\label{onest}\ea
where $h$ is the conformal dimension of the operator
${\cal{O}}_{1,2}$.

 At first, one may think they are decoupled because the
CQM1 and CQM2 are disconnected. However, as the non-vanishing
two point functions show, AdS/CFT predicts they are actually
correlated. A similar puzzle has been raised in \cite{MM} in the
context of AdS wormhole. Indeed the following discussion is closely
related to the holographic computation of entanglement entropy in
AdS wormholes \cite{HRT}.

In this paper we would like to claim that CQM1 and CQM2 are
actually quantum mechanically entangled with each other and that
this is the reason why we get the non-vanishing correlators. To show
that the two CFTs are entangled, we need to compute the entanglement
entropy and to check that it is non-zero. Below we would like to
calculate the entanglement entropy holographically.

The holographic formula (\ref{ARE}) is expected to be true in
general AdS space. If we apply it to our AdS$_2$ setup (i.e. d=0 in
(\ref{ARE})), we naturally find \be
S_{ent}=\f{\mbox{Area}(\g_A)}{4G^{(2)}_N}=\f{1}{G^{(2)}_N}
\label{entadst}. \ee This is because the minimal surface now becomes
a point. Below we will give a clearer derivation of (\ref{entadst})
based on the AdS/CFT.

The Hilbert spaces of CQM$1$ and CQM$2$ are denoted by $H_1$ and
$H_2$. The total Hilbert space looks like $H_{tot}=H_1\otimes H_2$.
We define the reduced density matrix from the total density matrix
$\rho_{tot}$ \be \rho_1=\mbox{Tr}_{H_2}\rho_{tot}, \ee by tracing
over the Hilbert space $H_2$. This is the density matrix for an
observer who is blind to CQM$2$. It is natural to assume that
$\rho_{tot}$ is the one for a pure state.

\begin{figure}[htbp]
\begin{center}
  \hspace*{0.5cm}
  \includegraphics[height=4.5cm]{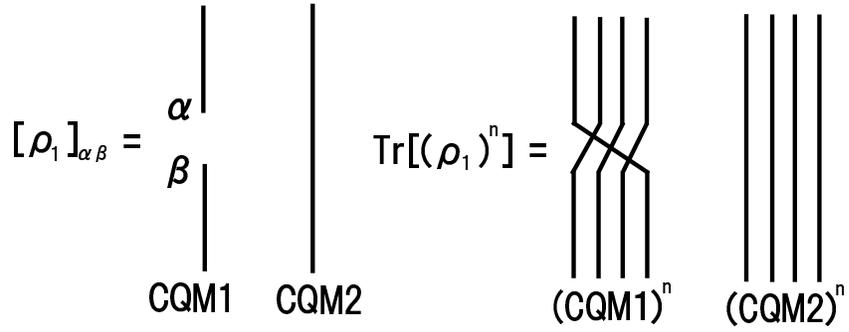}
  \caption{The calculation of reduced density matrix $\rho_1$}\label{qft}
\end{center}
\end{figure}

The entanglement entropy for CQM$1$, when we assume that the
opposite part CQM$2$ is invisible for the observer in CQM$1$, is
defined by \be S_{ent}=\mbox{Tr} [-\rho_1 \log \rho_1]. \ee We can
 obtain this by first computing Tr$(\rho_1)^n$, taking the
derivative w.r.t.$~n$ and finally setting $n=1$. In the path
integral formalism of the quantum mechanics, $\rho_1$ and
Tr$(\rho_1)^n$ are computed as in Figure \ref{qft} (we perform the
path-integral along the thick lines. $\ap$ and $\beta$ are the
boundary conditions.).

\begin{figure}
\begin{center}
  \includegraphics[height=6cm]{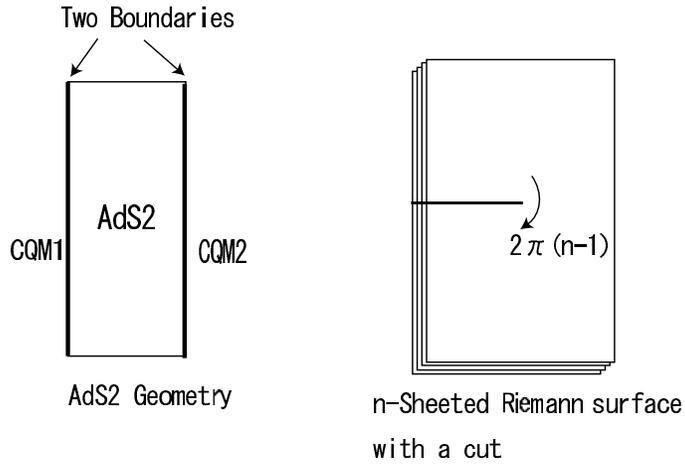}
  \caption{The geometry of AdS$_2$ [Left] and the 2D spacetime which is dual to the
  computation Tr$(\rho_1)^n$ [Right].}\label{adstwo}
\end{center}
\end{figure}

By using the bulk-boundary relation of AdS/CFT \cite{GKPW}, we can
compute the entanglement entropy holographically as in the right of
the Figure \ref{adstwo}. The dual geometry is the $n$-sheeted
Riemann surface \cite{RT}. Though our derivation below is along the
line with the argument in \cite{Fu} for $AdS_{d\geq 3}$ which proves
the claim in \cite{RT} via the bulk to boundary relation
\cite{GKPW}, our example is more non-trivial as it includes two
boundaries.
 Also it is closely
related\footnote{Notice that in these arguments the authors consider
the entanglement entropy for the {\it total spacetime} of
non-extremal black holes, while in our argument we consider the
entanglement entropy for {\it the boundary of} the extremal black
hole geometry.} to the conical defect argument of black hole entropy
(see e.g. \cite{Ja,FS}).

 Here we are considering an
Euclidean metric. The cut should end on a certain point in the bulk
because there should not be any cut on the opposite boundary, which
is first traced out. Notice that the presence of two boundaries in
AdS$_2$ plays a crucial role in this holographic computation. We
would get the vanishing entropy if we were to start with the
spacetime which has a single boundary such as the Poincare metric of
AdS$_2$.

Now we remember the Einstein-Hilbert action in the Euclidean space
\be S_{EH}=-\f{1}{16\pi G^{(2)}_N}\int dx^2 \s{g}(R+\Lambda). \ee
The cosmological constant $\Lambda$ is not important since it is
extensive and it will vanish in the end of the entropy computation.
In the $n$-sheeted geometry we find $S_{EH}=\f{n-1}{4G^{(2)}_N}$ in
the Euclidean formalism because the curvature behaves like a delta
function $R=4\pi (1-n)\delta^2(x)$ (see e.g.\cite{Fu,FS}). The
entanglement entropy is obtained as follows \be S_{ent}=-\f{\de}{\de
n}\log (e^{-S_{EH}+nS^{(0)}_{EH}})|_{n=1}=\f{1}{4G^{(2)}_N}, \ee
where $S^{(0)}_{EH}$ is the value of Einstein-Hilbert action of a
single-sheet in the absence of the cut (or negative deficit angle).

Finally, it is trivial to see that \be S_{ent}=S_{BH},
\label{entbb}\ee because $\frac{1}{G^{(2)}_N}=\frac{4\pi
r_+^2}{G^{(4)}_{N}}$. This means that the entanglement between
CQM$1$ and CQM$2$ is precisely the source of the 4D (near) extremal
black hole entropy. The same argument can be applied to any $d$
dimensional black holes or black rings whose horizons are of the
form AdS$_2\times M_{d-2}$, where $M_{d-2}$ is a compact manifold
such as $S^{d-2}$.

Recently, it has been shown that extremal (rotating) black holes
always have the $SO(2,1)$ symmetry in the near horizon limit
\cite{KLR,AGJST,AY}. For example, the near horizon geometry of a four
dimensional extremal Kerr black hole is given by a warped product of
AdS$_2$ and a two dimensional manifold \cite{BaHo}. Our argument in
this subsection can be applied to such a warped AdS$_2$ case.

\subsection{Higher Derivative Corrections}

\hspace{5mm} Moreover, we can take curvature corrections into
account. We assume that the near horizon geometry is of the form
AdS$_2 \times S^{d-2}$. Even though we start with the Lagrangian
${\cal L}$ that includes the curvature tensor $R_{\mu\nu\rho\sigma}$
and their covariant derivatives, we can neglect the covariant
derivative of curvature tensors because the near horizon geometry
has the constant curvature. In this case, the black hole entropy
with the curvature corrections is given by the Wald's formula
\cite{Wa,JKM} \be S_{bh}=-2\pi \int_{{\cal H}}\s{h}\f{\de
{\cal{L}}}{\de R_{\mu\nu\rho\sigma}}\ep_{\mu\nu}\ep_{\rho\sigma},
\ee where $\ep_{\mu\nu}=\xi_\mu\eta_\nu-\xi_\nu\eta_\mu$ by using
the Killing vector $\xi_\mu$ of the Killing horizon and its normal
$\eta_\nu$, normalized such that $\xi\cdot \eta=1$; ${\cal H}$
represents the horizon and $h$ is the metric on it.

For example, in the ordinary Einstein-Hilbert action $I=-\f{1}{16\pi
G_N}\int dx^{d}\s{g}R=-\int dx^d \sqrt{g}{\cal{L}}$, we reproduce the
standard result \be
S=\f{A_H}{8G_N}\ep_{\mu\nu}\ep_{\rho\sigma}g^{\mu\rho}g^{\nu\sigma}
=\f{A_H}{4G_N}, \ee where $A_H$ is the horizon area.

Now we would like to compare the Wald entropy with the entanglement
entropy computed holographically via AdS$_2$/CFT$_1$. We consider
the $n$-sheeted AdS$_2$ (times the same $S^{d-2}$), where the
Riemann tensor behaves as follows \cite{FS} \be
R_{abcd}=R_{abcd}^{(0)}+2\pi(1-n)\cdot
(g_{ac}g_{bd}-g_{ad}g_{bc})\cdot \delta_H\ . \ee Here $\delta_H$ is
the delta function localized at the (codimension two) horizon;
$R_{abcd}^{(0)}$ represents the constant curvature contribution from
the cosmological constant. $a,b$ run the coordinate in the AdS$_2$.
Notice also that if we employ the relation
$g_{ab}=\xi_a\eta_b+\xi_b\eta_a$, we obtain \be
\ep_{ab}\ep_{cd}=-(g_{ac}g_{bd}-g_{ad}g_{bc}). \ee

Now we consider the perturbative expansions of the Lagrangian with
respect to the (delta functional) deviation of $R_{abcd}$ from
$R_{abcd}^{(0)}$. Then the quadratic and higher order terms do not
contribute since $\lim_{n\to 1}\f{d}{dn}(1-n)^d=0$ for $d\geq 2$.
Therefore, we can find \ba I_n=-\log Z_n&=&-\int dx^d\s{-g}\f{\de
{\cal{L}}}{\de R_{abcd}} (-\ep_{ab}\ep_{cd})\cdot 2\pi
(1-n)\delta_H, \no &=&2\pi (1-n)\int_H \s{h}\f{\de {\cal{L}}}{\de
R_{abcd}}\ep_{ab}\ep_{cd}. \ea Thus this agrees with the Wald's
formula \be S_{ent}=-\f{\de}{\de n}\log Z_n\Bigl |_{n=1}=-2\pi\int_H\s{h}\f{\de
{\cal{L}}}{\de
R_{\mu\nu\rho\sigma}}\ep_{\mu\nu}\ep_{\rho\sigma}=S_{bh}. \ee

\subsection{Towards Holography in Flat Spacetime}

For general non-extremal black holes, \be
ds^2=-\frac{(r-r_+)(r-r_-)}{r^2}dt^2+\frac{r^2}{(r-r_+)(r-r_-)}dr^2+r^2
d\Omega^2, \ee we obtain a Rindler space in the near horizon limit
$r\to r_+~(>r_-)$. The global extension of the Rindler space is
clearly the two dimensional Minkowski spacetime $R^{1,1}$. Thus we
cannot relate it to the AdS/CFT correspondence.

If we associate two quantum mechanical systems, however, to two
time-like curves situated at the left and right side of $R^{1,1}$,
then we can obtain the same equality as (\ref{entbb}). This suggests
that the flat Minkowski spacetime may admit its holographic
description. It also has a natural higher dimensional extension by
expressing the $R^{1,d}$ metric as $ds^2=dr^2+r^2ds^2_{dS_{d}}$,
where $ds^2_{dS_{d}}$ is the metric of the $d$ dimensional de-Sitter
space.

\section{The AdS$_2/$CFT$_1$ Duality from 5D Near Extremal Black Holes}

\hspace{5mm} In the previous section we have argued that the black
hole entropy of (near) extremal black hole whose near horizon
geometry includes a AdS$_2$ factor, is equal to the entanglement
entropy of two dual CQMs, including quantum corrections. We
confirmed this claim by assuming the AdS$_2/$CFT$_1$. To obtain a
complete proof, we need to explicitly present a general construction
of the entangled pair of CQMs. In the 4D black hole cases, this is
not straightforward because the CQM dual to AdS$_2$ has not been
well understood at present.

Instead, in this section we would like to examine a concrete example
of AdS$_2/$CFT$_1$ which is obtained from the near extremal limit of
non-rotating 5D black holes. Equally we can regard this as a
dimensional reduction of AdS$_3/$CFT$_2$ as first pointed out in
\cite{St} since the near horizon geometry of 5D near extremal
black holes is a rotating BTZ black hole \cite{BTZ,BHTZ} (see also
\cite{MS}).

\subsection{Near Extremal BTZ Black Hole from 5D Black Holes}
\hspace{5mm} Consider a 5D black hole which is obtained from the
type IIB background with $Q_1$ D1-branes and $Q_5$ D5-branes wrapped
on $T^4\times S^1$ with Kaluza-Klein momentum $N$ in the $S^1$
direction. In the near horizon limit, the metric becomes
\cite{CaMa}\be \f{ds^2}{\al}=
\f{U^2}{l^2}(-dt^2+(dx^5)^2)+\f{U_0^2}{l^2}(\cosh\sigma
dt+\sinh\sigma
dx^5)^2+\f{l^2}{U^2-U_0^2}dU^2+l^2(d\Omega_3)^2+\s{\f{Q_1}{vQ_5}}dx_i^2.\ee
Via a coordinate transformation we can show that this geometry is
equivalent to \cite{MS} \be (\mbox{BTZ black hole})_3\times
S^3\times T^4. \ee The metric of the rotating BTZ black hole metric
\cite{BTZ,BHTZ} is given by (\ref{metnon}). The explicit coordinate
transformation is given by \be t \to bt,\quad x^5 \to bR\phi,\quad
(U^2 + U_0^2\sinh ^2\sigma) \to \frac{r^2}{b^2}, \qquad \text{for}~~
\forall b\ee and the new parameters are defined as
$R=l,~r_+=bU_0\cosh\sigma,~r_-=bU_0\sinh\sigma$. We can take $\phi
\sim \phi + 2\pi$ if we choose $b=R_5/R$, where $R_5$ is the radius
of $x_5$.

This BTZ geometry (\ref{metnon}) can also be obtained from a
Lorentzian orbifold of the pure AdS$_3$ space \be\label{eq:Poincare}
ds^2=R^2\f{dy^2+dw_+dw_-}{y^2}. \ee They are related by the
coordinate transformation  \be
w_{\pm}=\s{\f{r^2-r^2_+}{r^2-r^2_-}}e^{\f{r_+\pm
r_-}{R}\left(\pm\f{t}{R}+\phi\right)},\ \ \ \
y=\s{\f{r^2_+-r^2_-}{r^2-r^2_-}}e^{\f{r_+}{R}\phi+\f{r_-}{R^2}t}.
\ee The periodicity of $\phi$ (i.e. $\phi\sim \phi+2\pi$) leads to
the identification \be w_{+}\sim e^{4\pi^2 T_L}w_+, \ \ \ \
w_{-}\sim e^{4\pi^2 T_R}w_-,\ \ \ \ y\sim e^{2\pi^2(T_L+T_R)} y ,
\ee where $T_{L}=\f{r_++r_-}{2\pi R}$ and $T_{R}=\f{r_+-r_-}{2\pi
R}$ represent the left and right-moving temperature of the dual 2D
CFT. The central charge of dual CFT is given by
$c=\f{3R}{2G^{(3)}_N}$ and its density matrix looks
like \be \rho=e^{-\f{L_0}{T_L}-\f{\bar{L}_0}{T_R}}, \ee using the
left and right-moving energy $L_{0}$ and $\bar{L}_0$.

In the extremal case $r_+=r_-$, we need another coordinate
transformation defined by \be
w_{+}=\f{R}{2r_+}e^{\f{2r_+}{R}\left(\f{t}{R}+\phi\right)},\ \ \ \ \
w_-=\phi-\f{t}{R}-\f{Rr_+}{r^2-r_+^2},\ \ \ \ \
y=\f{R}{\s{r^2-r_+^2}}e^{\f{r_+}{R}(\f{t}{R}+\phi)}. \ee The
periodicity of $\phi$ is equivalent to \be\label{eq:BTZident}
w_+\sim e^{4\pi^2 T_L}w_+,\ \ \ \ w_-\sim w_-+2\pi ,\ \ \ \ y\sim
e^{2\pi^2 T_L}y. \ee The thermal entropy of the dual CFT is given by
the standard formula $S_A=\f{\pi^2}{3}cT_L$ and this agrees with the
black hole entropy $S=2\pi\s{\f{cL_0}{6}}=2\pi\s{Q_1Q_5N}$, using
the thermodynamical relation $L_0=\f{\pi^2}{6}cT_L^2$.

\subsection{From Near Extremal BTZ to AdS$_2$}
\hspace{5mm} The near extremal 5D black hole is related to the near
extremal BTZ black hole $\f{r_+-r_-}{r_+}\ll 1$. In the dual CFT,
the left moving sector is far more excited compared with the
right-moving sector since $T_L\gg T_R$.

By considering the limit $r\to r_+$ of the BTZ metric
(\ref{metnon}), we define $u=r-r_+$ and assume $u\sim (r_+-r_-)\ll
r_+$.  In the end we find the simplified metric \be
ds^2=-\f{4u(u+r_+-r_-)}{R^2}dt^2+\f{R^2}{4u(u+r_+-r_-)}du^2+r_+^2(\f{dt}{R}+d
\phi)^2. \label{AdStb} \ee The 2D part of (\ref{AdStb}) is
equivalent to the 'AdS$_2$ black hole' defined in \cite{SS} \be
ds^2=-\f{u(u+4\pi Q^2 T_H)}{Q^2}dt^2+ \f{Q^2}{u(u+4\pi Q^2
T_H)}du^2, \ee where \be Q^2=\f{R^2}{4},\ \ \ \ T_H=\f{r_+-r_-}{\pi
R^2}. \ee  We can show that this space is equivalent to the pure
AdS$_2$ via a coordinate transformation \cite{SS}. Though the
temperature dependence disappears by this transformation, it
reflects the choice of different thermal vacua \cite{SS}. Thus the
3D background (\ref{AdStb}) is equivalent to AdS$_2\times S^1$.

In order to have a sensible interpretation in terms of
AdS$_2$/CFT$_1$, the geometry should include the boundary region of
the AdS$_2$ dual to the UV limit of CFT$_1$. This is given by the
region $u\gg R$. On the other hand, the approximation to get
(\ref{AdStb}) assumes the condition $u\ll r_+$. Thus we have to
require \be R\ll r_+.\label{ol} \ee This means that we cannot
neglect the excitation in the $S^1$ direction of the spacetime
AdS$_2\times S^1$. However, still we can perform the Kaluza-Klein
reduction and regard the theory as the one on AdS$_2$ with
infinitely many Kaluza-Klein modes.

The generators\footnote{Notice that we distinguish $l_{0,\pm 1}$
from the standard basis $L_{0,\pm 1}$ dual to the Virasoro
generators of 2D CFT. In our case, the unbroken generators of
$l_{0,\pm 1}$ and $\bar{l}_{0,\pm 1}$ are linear combinations of the
standard Virasoro generators.} $l_0,\ l_{\pm 1}$ and $\bar{l}_0,\
\bar{l}_{\pm 1}$ of the isometry $SO(2,2)=SL(2,R)_L\times SL(2,R)_R$
of the AdS$_3$ in the Poincare coordinate (\ref{eq:Poincare}) are
given by \be l_{-1}=-\de_{w_+},\ \ \
l_0=-(w_+\de_{w_+}+\f{1}{2}y\de_y),\ \ \
l_1=-(w_+^2\de_{w_+}+w_+y\de_{y}-y^2\de_{w_-}), \label{expl} \ee and
their anti-holomorphic counterparts obtained by exchanging $w_{\pm}$
with $w_{\mp}$. For states dual to
 generic BTZ black holes, the two $SL(2,R)$ symmetries are both
 broken. However, if we take the limit $R\to 0$ (i.e.\ (\ref{ol}))
of the extremal
 BTZ $r_+=r_-$, we can keep $U(1)_L\times SL(2,R)_R$ (i.e.\
$l_0$ and $\bar{l}_{\pm1},\bar{l}_0$) unbroken as is clear from the
orbifold action (\ref{eq:BTZident}) on the expressions (\ref{expl}).
 The generator $U(1)_L$ is the translation in the $S^1$ direction
and the right-moving $SL(2,R)_R$ symmetry turns out to be
essentially the same as the isometry of the AdS$_2$ \cite{St}.

This analysis of the conformal symmetry reveals that the excitation
in the $S^1$ direction is related to the left-moving sector. Thus we
can regard this AdS$_3/$CFT$_2$ as a variant of AdS$_2/$CFT$_1$ by
treating the left-moving sector as an internal degree of freedom.
Notice that excitations in the left-moving sector do not shift the
value of the Hamiltonian for CFT$_1$ (i.e.\ $\bar{L}_0$). Thus the
conformal quantum mechanics dual to AdS$_2$ is essentially described
by the right-moving part of CFT$_2$.

This suggests a DLCQ interpretation of the dual CFT. In order to
properly normalize the metric (\ref{AdStb}) in the limit (\ref{ol}),
we are lead to define \be X^+=\f{r_+}{R}\left(\f{t}{R}+\phi\right),\
\ \ \ \ X^-=\f{R}{r_+}\left(\f{t}{R}-\phi\right).  \label{boost} \ee
Thus in this picture we can equivalently regard that the CFT$_2$ is
almost light-like compactified $X^+\sim X^++\f{2\pi r_+}{R}$ and
$X^-\sim X^-+\f{2\pi R}{r_+}$. In this description, it is easy to
confirm the unbroken $SL(2,R)$ symmetry because $w_-$ is scaled as
$\f{R}{r_+}w_-$ and gets insensitive under the orbifold action. Also
this rescaling shifts the energy scale we are looking at as
$(p_+,p_-)\to (\f{R}{r_+}p_+, \f{r_+}{R}p_-)$. This agrees with the
near extremal limit $L_0\sim \f{r_+}{R}\gg 1$ that we have been
assuming so far. Notice also that in this limit the time evolution
is equivalent to the one of the light-cone time $X^-$ and therefore
the right-moving energy $\bar{L}_0$ is treated as the Hamiltonian.

\subsection{Two Point Functions}
\hspace{5mm} In order to have a better understanding of the
AdS$_2/$CFT$_1$ interpretation of the near extremal BTZ black hole,
we would like to turn to the two point function computed
holographically following the bulk to boundary relation \cite{GKPW}.

The Feynman Green function of a scalar field in global AdS$_3$ is
given in \cite{IS} and also that in BTZ can be constructed by the
orbifold method. AdS$_3$ is defined as the three dimensional
hyperboloid $-x_0^2-x_1^2+x_2^2+x_3^2 = -R^2$ embedded in $R^{2,2}$
and its metric takes a form $ds^2 = -dx_0^2 - dx_1^2 + dx_2^2 +
dx_3^2$. In the global AdS$_3$, the Green function takes fairly
simple form like \be\label{eq:GreenGlobal} -iG_F(x, x') = \f{1}{4\pi
R}(z^2-1)^{-1/2}\left[
  z +(z^2-1)^{1/2}  \right]^{1-2h_+},
\ee where \beq
z &\equiv& 1 + R^{-2}\sigma (x, x') + i\epsilon, \nonumber\\
\sigma (x, x') &=& \f{1}{2}\eta_{\m\n}(x-x')^\m (x-x')^\n,\quad
\eta_{\m\n} = diag(-1,-1,1,1). \eeq If we define the coordinate \beq
x_0 &=& \f{y}{2}\left( 1+\f{1}{y^2}(R^2 + w_+w_-) \right), \nonumber\\
x_1 &=& \f{R}{2y}(w_+ - w_-),\nonumber\\
x_2 &=& \f{y}{2} \left( 1-\f{1}{y^2}(R^2 - w_+w_-) \right),
\nonumber\\
x_3 &=& \f{R}{2y}(w_+ + w_-), \eeq we obtain the Poincare coordinate
(\ref{eq:Poincare}). The parameter $z$ in the above coordinate
becomes \be\label{eq:zPoincare} z^{(Poincare)} = \f{1}{2yy'}\left[
y^2 + y'^2 + \D w_+ \D w_- \right] \ee and by substituting this to
(\ref{eq:GreenGlobal}), we obtain the Green function in the Poincare
coordinate.

Considering the images which come as a result of the orbifolding
procedure, the Green function in the rotating BTZ
becomes \beq -iG_{non-extBTZ}(x, x') &=& \f{1}{4\pi
  R}\sum_{n=-\infty}^{\infty}(z_n^2-1)^{-1/2}
\left[ z_n +(z_n^2-1)^{1/2}  \right]^{1-2h_+},\\
z_n(x, x') -i\epsilon &=& \f{1}{r_+^2 - r_-^2}\Bigg[
  \s{r^2-r_-^2}\s{r'^2-r_-^2}\cosh\left( \f{r_-}{R^2}\D t_n +
    \f{r_+}{R}\D \phi_n\right) \nonumber\\
&& \quad + ~~\s{r^2-r_+^2}\s{r'^2-r_+^2}\cosh\left( \f{r_+}{R^2}\D
t_n -
    \f{r_-}{R}\D \phi_n\right) \Bigg],
\eeq where \beq
  \D t_n = t - t',\qquad \D\phi_n = \phi - \phi' +2\pi n . \eeq

Now we would like to reduce the previous bulk-bulk Green functions
to the AdS$_2$ ones. Notice that the geodesic length $z_n$ can
always be taken to be very large since we can consider two points
near the boundary of AdS$_2$ owing to (\ref{ol}). Thus the Green
function looks like \be G\sim \f{1}{4\pi R}\sum_{n=-\infty}^\infty
(z_n)^{-2h_+}. \ee

Consider again the near extremal BTZ $\f{r_+-r_-}{r_+}\ll 1$ and
take the limit $u=r-r_+\gg R\sim r_+-r_-$. Then we obtain \ba
z_n&\sim& \f{\s{yy'}}{r_+^2-r_-^2} \left[\cosh\left(\f{r_-\Delta
t}{R^2}+\f{r_+\Delta \phi_n}{R}\right)-\cosh\left(\f{r_+\Delta
t}{R^2}+\f{r_-\Delta \phi_n}{R}\right)\right]\no &=&\!\!
2\f{\s{yy'}}{r_+^2-r_-^2}\sinh\!\left(\!\f{(r_++r_-)
}{2R}\left(\f{\Delta t}{R}+\Delta
\phi_n\right)\!\!\right)\!\sinh\left(\f{(r_+-r_-)
}{2R}\left(\f{\Delta t}{R}-\Delta \phi_n\right)\!\!\right).\ea

In this case the holographic two point function in the AdS$_2$ limit
looks like (below we omit numerical constants) \be \la
O(t,\phi)O(0,0)\lb =\sum_{n}\left[\sinh\left(\f{(r_++r_-)
}{2R}\left(\f{\Delta t}{R}+\Delta
\phi_n\right)\right)\sinh\left(\f{(r_+-r_-) }{2R}\left(\f{\Delta
t}{R}-\Delta \phi_n\right)\right)\right]^{-2h_+}. \ee This takes the
same expression as the one of holographic two point function of
CFT$_2$ \cite{Kraus,KOS}. In the DLCQ coordinate, this is rewritten
as follows \be \la O(X^{+},X^-)O(0,0)\lb =\sum_{n} \left[\sinh
\left(X^++\f{2\pi r_+}{R}n\right) \sinh \left(\f{(r_+-r_-)r_+}{2
R^2}X^-+\pi n(r_+-r_-)\right)\right]^{-2h_+}. \label{cortt}\ee

In the DLCQ coordinate, we treat $X^\pm$ as the basic coordinates and
thus in the scaling (\ref{ol}) we can set $n=0$ in the above
summation. Then the left and right-moving sector are decoupled as
expected. Notice that the coordinate in the $S^1$ direction is
$X^+$.

To interpret the near horizon limit of near extremal BTZ (i.e.\
AdS$_2\times S^1$) from the viewpoint of the AdS$_2/$CFT$_1$, we
need to regard the left-moving sector dual to the $S^1$ part as an
internal degree of freedom as we have explained before. This allows
us to treat $X^+$ as a label of internal quantum number. Thus we can
extract the two point function of CFT$_1$ from (\ref{cortt}) as
follows \be \la O(t)O(0)\lb=[\sinh(\pi T_H t)]^{-2h_+}. \ee Here we
have employed the relation $X^-\sim \f{2}{r_+}t$, which is obtained
from the infinite boost (\ref{boost}).  This behavior agrees with
the result for the thermal state in AdS$_2$ \cite{SS}. Especially,
in the extremal limit $T_H\to 0$ we find \be \la
O(t)O(0)\lb=t^{-2h_+}, \ee as expected. In this way we have
confirmed that we can regard the AdS$_3/$CFT$_2$ correspondence for
the near extremal BTZ black hole equally as the AdS$_2/$CFT$_1$ with
infinitely many internal degrees of freedom.

\subsection{Quantum Entanglement and Black Hole Entropy}
\hspace{5mm} As we have explained, the AdS$_3/$CFT$_2$
correspondence for the near extremal black holes can also be
regarded as a AdS$_2/$CFT$_1$ by taking the near horizon limit of
the near extremal BTZ black hole. Essentially, the CFT$_1$ i.e. the
conformal quantum mechanics is described by the right-moving sector
of the original CFT$_2$ by treating the left-moving one as an
internal degree of freedom tensored with the right-moving sector.
When we consider the excitation in the AdS$_2$ spacetime with the
$S^1$ sector untouched, the left-moving sector will always stay at
$L_0=N$, where $N$ is the quantized momentum in the original 5D
black hole description.

Usually, the CFT dual of the rotating BTZ black hole is interpreted
as a thermal state. Equally we can interpret this as an entangled
state in two copies of the same CFT \cite{ME} \be
|\Psi\lb=\f{1}{\s{Z_0}}\sum_{n_L,n_R}e^{-\beta_L L_0/2-\beta_R
\bar{L}_0/2}\left(|n_L\lb_L \otimes |n_R\lb_R \right)_{CFT1}\otimes
\left(|n_L\lb_L \otimes |n_R\lb_R \right)_{CFT2}, \ee where
$Z_0=\sum_{n_L,n_R}e^{-\beta_L L_0-\beta_R \bar{L}_0}$ is the
partition function of the 2D CFT. In the gravity side, they are
geometrically understood as the CFTs living on the two disconnected
boundaries of the BTZ spacetime.

To describe near extremal BTZ black holes, we keep $\beta_L$ finite
and $\beta_R$ very large. In the near horizon limit $r\to r_+$, two
boundaries of BTZ descend to the direct product of the two
boundaries of AdS$_2$ times the circle $S^1$. We denote the states
with $L_0=N$ by $|k\lb$ $(k=1,2,\ddd,d(N))$. The number $d(N)$ of
such states is very large $d(N)\sim e^{2\pi\s{Q_1Q_5N}}$. Then the
quantum state looks like \be\label{eq:QSCQM} |\Psi\lb
=\f{1}{\s{d(N)}}\sum_{n}\sum_{k=1}^{d(N)}e^{-\beta
E_n/2}\left(|k\lb_L \otimes |n\lb_R \right)_{CFT1}\otimes
\left(|k\lb_L \otimes |n\lb_R \right)_{CFT2}, \ee where $E_n=\la
n|\bar{L}_0|n\lb$ is the energy of the CQM.

Consider the zero temperature limit $\beta=\infty$. Then the
right-moving sector has a single ground state $|0\lb$. The reduced
density matrix of CQM1 $\rho_1$, which is obtained by tracing over
CQM2, now becomes \be \rho_{1}=\f{1}{\s{d(N)}}\sum_{k=1}^{d(N)}
|k\lb \la k |_{CQM1}\ \ \ \ , \label{den} \ee  where
$|k\lb_{CQM1}=|k\lb_L\otimes |0\lb_R$. This leads to the following
entanglement entropy \be S_1=\mbox{Tr}[-\rho_1\log\rho_1]=\log d(N)
= 2\pi\s{Q_1Q_5N}. \ee This clearly agrees with the familiar
microscopic counting of BPS states and thus is equal to the black
hole entropy \cite{CaMa}. We can also confirm that it agrees with
the entanglement entropy calculated holographically for the near
horizon geometry AdS$_2\times S^1 \times S^3\times T^4$ of 5D (near)
extremal black holes. In this way, we have shown that the
AdS$_2/$CFT$_1$ description correctly reproduces the black hole
entropy of (near) extremal 5D black holes.

We would like to stress that the density matrix (\ref{den}) shows
that the two quantum mechanics are maximally entangled. In general
it is possible to find a quantum state with a smaller value of
entanglement entropy $S_1<\log d(N)$ even if the number of
degeneracy is $d(N)$. However, the entropy of extremal black holes
known so far has always been explained by assuming maximally
entangled states.

\subsection{Subtlety of the Extremal Limit}

\hspace{5mm} In this section we have mostly treated the extremal BTZ
black holes as a limit of non-extremal ones, instead of starting
with the extremal ones themselves. This is because the extremal
limit looks sometimes subtle. This subtlety of defining extremal
black hole entropy has been noticed for a long time \cite{HHR}.

\begin{figure}
\begin{center}
  \includegraphics[height=6cm]{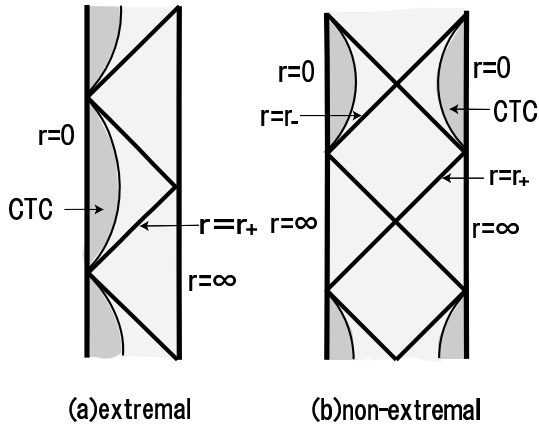}\\
  \caption{Penrose diagrams of the extremal and non-extremal
    BTZ black hole. There is a closed timelike curve in the shaded region.
}
\label{penrose}
\end{center}
\end{figure}

First of all, this subtlety is noticed from the different forms of
Penrose diagrams (Figure \ref{penrose}) \cite{BHTZ}. In both
extremal and non-extremal case, there are two boundaries in the
Penrose diagram. Thus one may think that they should be interpreted
as the two entangled CFTs. However, in the extremal case one of the
two boundaries always includes the closed time like curve (Figure
\ref{penrose}(a)), while in the non-extremal case not (Figure
\ref{penrose}(b)). As far as we consider the non-extremal case, we
can find the same boundary structure in the opposite boundary (as in
Figure \ref{penrose}(b)) and thus we can apply the
interpretation\footnote{It is often claimed that we cannot extend
the rotating black hole spacetime beyond the inner horizon
\cite{ME,LR,BaLe}. Our derivation of black hole entropy from the
holographic entanglement entropy done in section 2 is still fine
even if we take this restriction into account.} of two entangled
CFTs \cite{ME,Kraus,KOS,LR,BaLe,MaYa}.

In the extremal case, we find only one boundary which has the
sensible property with the CFT dual. Therefore one may worry that
the entangled interpretation is confusing in the strictly extremal
case. On the other hand, most of physical quantities of extremal
black holes such as two-point functions are obtained smoothly by
taking the extremal limit $r_+\to r_-$ of those of the non-extremal
ones. Therefore, if we apply the previous analysis in the
non-extremal case to the extremal case, we will get the same
conclusion; the CFT dual to the extremal case is described by the
entangled states. Refer also to the argument in \cite{MaYa} for an
interesting candidate of a geometrical interpretation of these
entangled pairs via AdS$_3/$CFT$_2$.

Even though we cannot completely resolve the mentioned conflict with
the global geometry, the holographic consideration leading to
(\ref{den}) via AdS$_3/$CFT$_2$, tells us that the entangled
interpretation is still correct even for strictly extremal black
holes. Also notice that in the near horizon limit $r\sim r_+$, we do
not have to worry about this problem. This is because the near
horizon geometry of the extremal case has no closed time-like curve
and two regular boundary CFTs are recovered in this limit. It will
be an interesting future problem to explore this point.

\section{Finite Size Corrections of Entanglement Entropy at Finite Temperature}

\hspace{5mm} In this section we compute the entanglement entropy of
a 2D free Dirac fermion at finite temperature when the spatial
direction is compactified (to unit radius). This is the first
analytical result of the entanglement entropy for a finite size 2D
CFT at finite temperature. In the case of either infinite size or
zero temperature, the expression of entanglement entropy becomes
very simple and takes the form of the central charge $c$ times a
universal function as found in \cite{HLW,Cardy}. However, in our
case, the entanglement entropy depends more sensitively on the
theory we consider.

In section 2, we have seen that the relation (\ref{relatione}) is
very important for the understanding of BTZ black hole entropy in
AdS$_3$/CFT$_2$. This important relation between thermal entropy and
entanglement entropy can only be explicitly shown in a finite size
system. Indeed the behavior of the entanglement entropy agrees with
what we expect from the geometric picture obtained from the AdS/CFT
explained in section 2. This supports our claim that the black hole
entropy is interpreted as the entanglement entropy in the dual CFT.

\subsection{Two Point Function of a Compactified Boson}

\hspace{5mm} To make calculations simple, we consider the
entanglement entropy of a free Dirac fermion $\psi$. This fermion is
bosonized into a scalar field $\vp$ with the unit radius $R=1$ as
$\psi=e^{i\vp}$. We assume the Euclidean 2D theory on a torus
defined by $z\sim z+1$ and $z\sim z+\tau$ since we are interested in
a finite temperature theory with a finite size. In particular, when
the period $\tau$ is pure imaginary $\tau=i\beta$, the theory is at
the temperature $\beta^{-1}$ and its spacial size is $1$.

The primary operator $O_{(n,w)}$ denotes the one with the momentum
$n$ and the winding $w$ such that the chiral dimension becomes
$\Delta_{n,w}=\f{1}{2}\left(\f{n}{R}+\f{wR}{2}\right)^2$ and
$\bar{\Delta}_{n,w}=\f{1}{2}\left(\f{n}{R}-\f{wR}{2}\right)^2$.

Their two point functions are given by (see e.g. section 12 in
\cite{DMS}) \ba &&\la O_{(n,w)}(z,\bar{z})\ O_{(-n,-w)}(0,0)\lb =
\no &&\ \ \
\left(\f{2\pi\eta(\tau)^3}{\theta_1\left(z|\tau\right)}\right)^{2\Delta_{n,w}}\cdot
\overline{\left(\f{2\pi\eta(\tau)^3}{\theta_1
\left(z|\tau\right)}\right)}^{2\bar{\Delta}_{n,w}}\cdot
\f{\sum_{m,l}q^{\Delta_{m,l}}\bar{q}^{\bar{\Delta}_{m,l}}e^{4\pi
i(\ap_{n,w}\ap_{m,l}z-\bar{\ap}_{n,w}\bar{\ap}_{m,l}\bar{z})}}
{\sum_{m,l}q^{\Delta_{m,l}}\bar{q}^{\bar{\Delta}_{m,l}}}, \label{twop} \ea
where $\ap_{n,w}\equiv \f{1}{\s{2}}\left(\f{n}{R}+\f{Rw}{2}\right)$
and $\bar{\ap}_{n,w}\equiv
\f{1}{\s{2}}\left(\f{n}{R}-\f{Rw}{2}\right)$.

In particular, we are interested in a Dirac fermion, which is
equivalent to the real boson at the radius $R=1$. For example, the
one-loop partition function $Z_{bos}(R)$ is transformed as follows
\ba Z_{bos}(R=1)\cdot
|\eta(\tau)|^2&=&\sum_{n,w}q^{\f{(n+w/2)^2}{2}}\bar{q}^{\f{(n-w/2)^2}{2}}
\no &=&
\f{|\theta_2(0|\tau)|^2+|\theta_3(0|\tau)|^2+|\theta_4(0|\tau)|^2}{2}.\ea
In this way the free boson partition function is decomposed into the
four sectors $(R,NS)$, $(NS,NS)$, $(NS,R)$ and $(R,R)$, each
corresponds to $\nu=2,3,4,1$ of the theta function $\theta_{\nu}$ as
usual.

\subsection{Calculating Entanglement Entropy}
\hspace{5mm} In general, to compute the entanglement entropy, we
first divide the total system into two subsystems $A$ and $B$. In
our setup, we define $A$ (or $B$) to be an interval with length $L$
(or $1-L$) at a specific time. Next, we compute Tr$(\rho_A)^N$,
where $\rho_A$ is the reduced density matrix obtained by taking a
trace of the density matrix $\rho$ over the subsystem $B$ i.e.\
$\rho_A=\mbox{Tr}_B \rho$. This is usually possible by assuming $N$
is an positive integer. Then we analytically continue with respect
to $N$. Finally we take the derivative of $N$ and obtain the
entanglement entropy $S_A$ of the subsystem $A$ \be S_A=-\f{\de}{\de
N}\log \mbox{Tr}(\rho_A)^N\Bigl|_{N=1}. \label{entdef}\ee

 We can calculate Tr$(\rho_A)^N$ by employing the following formula
 which relates it to a product of two point functions of twisted
 operators
 \cite{Casini,RT}
 \be \mbox{Tr}(\rho_A)^N=\prod_{k=-\f{N-1}{2}}^{\f{N-1}{2}}
\la \sigma_{k}(z,\bar{z})\sigma_{-k}(0,0) \lb, \label{entdeff} \ee
with the understanding of $z=L$.

We identify the twist operator $\sigma_{k}$ with the operator
$O_{(0,\f{k}{N})}$ which has the fractional winding number
$w=\f{2k}{N}$ so that the fermion $\psi=e^{i\vp}$ picks up the phase
$e^{\pm \f{2\pi i}{N}}$ if it goes around the two end points $0$ and
$z$. By setting $z=L$, we find the extra phase becomes \be e^{4\pi
i(\ap_{n,w}\ap_{m,l}z-\bar{\ap}_{n,w}\bar{\ap}_{m,l}\bar{z})}
=e^{4\pi i\f{mk}{N}L}
.\ee  Thus the two
point function (\ref{twop}) in the $\nu=2,3,4$ sector of the fermion
becomes \be \la \sigma_{k}(z,\bar{z})\sigma_{-k}(0,0)
\lb_{\nu}=\left|\f{2\pi\eta(\tau)^3}{\theta_1\left(L|\tau\right)}\right|^{4\Delta_{k}}\cdot
\f{|\theta_\nu\left(\f{kL}{N}|\tau\right)|^2}{|\theta_\nu(0|\tau)|^2},
\label{cort} \ee where $\Delta_{k}=\f{k^2}{2N^2}$. Below we assume
that $\tau=i\beta$ is pure imaginary except in section 4.7.

Now the entanglement entropy can be found by applying (\ref{entdef})
and (\ref{entdeff}) to (\ref{cort}). To make the presentation
simpler, we divide the entropy into two parts \be S_A=S_1+S_2, \ee
where $S_1$ is the one from the first factor in the right-hand side
of (\ref{cort}), while $S_2$ is from the second factor.

It is easy to calculate $S_1$ since the expression depends on $N$
only via the conformal dimension $\sum_{k}\Delta_k=\f{c}{24}(N-1/N)$
(in our model the central charge is given by $c=1$). We obtain \be
S_1=\f{c}{3}\log \left|\f{\theta_1\left(L|\tau
\right)}{2\pi\eta(\tau)^3}\right|. \ee The exact expression suitable
for the low temperature expansion is given by \be S_1=\f{c}{3}\log
\left|\f{1}{\pi}\cdot\sin(\pi L)\prod_{m=1}^\infty \f{(1-e^{2\pi
iL}q^m)(1-e^{-2\pi iL}q^m)}{(1-q^m)^2} \right|,\ee where $q=e^{-2\pi
\beta}$. The expression of high temperature expansion is obtained
from the modular transformation as follows \be S_1=\f{c}{3}\log
\left|\f{\beta}{\pi}\cdot e^{-\f{\pi
L^2}{\beta}}\cdot\sinh\left(\f{\pi L}{\beta}\right)
\prod_{m=1}^\infty \f{(1-e^{2\pi L/\beta}\ti{q}^m)(1-e^{-2\pi
L/\beta}\ti{q}^m)}{(1-\ti{q}^m)^2} \right|, \ee where
$\ti{q}=e^{-\f{2\pi}{\beta}}$. Notice that this contribution
satisfies \be S_1(L)=S_1(1-L)=S_1(1+L). \ee

Secondly, $S_2$ is given by \be S_2=-\f{\de}{\de
N}\sum_{k=-\f{N-1}{2}}^{\f{N-1}{2}}\log
\left|\f{\theta_\nu\left(\f{kL}{N}|\tau\right)}{\theta_\nu(0|\tau)}\right|^2
\Biggr|_{N=1}. \label{esutwo}\ee In order to perform an analytical
continuation with respect to $N$ we need to complete the summation
of $k$. This can be done by expanding the logarithm in
(\ref{esutwo}) explicitly by employing the standard formula
$\log(1+x)=\sum_{l=1}^\infty \f{(-1)^{l-1}}{l}x^l$ as we will see in
the next subsection.

\subsection{High Temperature Expansion}
\hspace{5mm} We first restrict to the special case $\nu=3$, i.e. the
NS sector for simplicity. We will come back to other spin structures
in section \ref{otherspinstructure}.

Let us evaluate $S_2$ in the high temperature expansion. In order to
get the high temperature expansion, we need to perform the modular
transformation $\tau\to -\f{1}{\tau}$ \be
\f{\theta_3(z|\tau)}{\theta_3(0|\tau)}=e^{-i\pi z^2/\tau}\cdot
\f{\theta_3(\f{z}{\tau}|-\f{1}{\tau})}{\theta_3(0|-\f{1}{\tau})}.
\ee Then we obtain \be S_2=-\f{\de}{\de
N}\sum_{k=-\f{N-1}{2}}^{\f{N-1}{2}}\left[-2\pi \f{k^2L^2}{\beta
N^2}\right]\Bigl|_{N=1}+\ti{S}_2 =
\f{\pi}{3}\cdot\f{L^2}{\beta}+\ti{S}_2, \ee where the part
$\ti{S}_2$ is found to be \ba \ti{S}_2&=&-2\f{\de}{\de
N}\sum_{k=-\f{N-1}{2}}^{\f{N-1}{2}}\sum_{m=1}^\infty
\log\left[\f{(1+e^{2\pi\f{kL}{N\beta}}e^{-2\pi
(m-1/2)/\beta})(1+e^{-2\pi\f{kL}{N\beta}}e^{-2\pi
(m-1/2)/\beta})}{(1+e^{-2\pi
(m-1/2)/\beta})^2}\right]\Biggr|_{N=1}\no &=& -8\f{\de}{\de
N}\sum_{k=-\f{N-1}{2}}^{\f{N-1}{2}}\sum_{m=1}^\infty\sum_{l=1}^\infty
\f{(-1)^{l-1}}{l}\cdot \sinh^2\left(\f{\pi kLl}{N\beta}\right)
e^{-2\pi(m-1/2)\f{l}{\beta}}\Bigl|_{N=1}\no
&=&- \sum_{l=1}^\infty \f{(-1)^{l-1}}{l}\left[ \f{2\pi
Ll}{\beta}\coth\left(\f{\pi
Ll}{\beta}\right)-2\right]\f{1}{\sinh\left(\f{\pi l}{\beta}\right)}.
\label{eq:EEnu=3}\ea In this calculation we have employed the
following formula \be \f{\de}{\de
N}\!\!\!\sum_{k=-\f{N-1}{2}}^{\f{N-1}{2}}\sinh^2\left(\f{\ap
k}{N}\right)\Biggr|_{N=1}=\f{\de}{\de N}\!\!\! \left[-\f{N}{2}+
\f{e^{\f{(1-N)\ap}{N}}-e^{\f{(N+1)\ap}{N}}}{2(1-e^{\f{2\ap}{N}})}\right]\Biggl|_{N=1}
=-\f{1}{2}+\f{\ap}{2}\coth \ap.\ee

\begin{figure}[htbp]
\begin{center}
  \includegraphics[height=3cm]{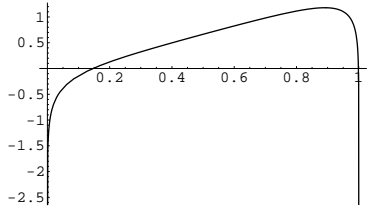}\\
  \caption{The entanglement entropy as a function of $L$ when
  $\beta=0.6$. We get rid of the divergence due to the cut off by setting
  $a=\f{1}{2\pi}$. }\label{enpl}
\end{center}
\end{figure}

In summary, the total expression of $S_A$ in the high temperature
expansion becomes \ba S_{A}&=&\f{1}{3}\log\left[\f{\beta}{\pi
 a}\sinh\left(\f{\pi
L}{\beta}\right)\right]+\f{1}{3}\sum_{m=1}^\infty
\log\left[\f{(1-e^{2\pi \f{L}{\beta}}e^{-2\pi \f{m}{\beta}
})(1-e^{-2\pi \f{L}{\beta}}e^{-2\pi \f{m}{\beta}})}{(1-e^{-2\pi
\f{m}{\beta}})^2}\right]\no  &&\quad + 2\sum_{l=1}^\infty
\f{(-1)^{l}}{l}\cdot\f{\f{\pi Ll}{\beta}\coth\left(\f{\pi
Ll}{\beta}\right)-1}{\sinh\left(\pi \f{l}{\beta}\right)}.
\label{toth}\ea In this final expression, we make the dependence on
the UV cut off $a$ explicit. We plotted the function (\ref{toth}) in
Figure \ref{enpl} by setting $a=\f{1}{2\pi}$ and $\beta=0.6$.

The first factor $\f{1}{3}\log\left[\f{\beta}{\pi
 a}\sinh\left(\f{\pi
L}{\beta}\right)\right]$ reproduces the known result in the infinite
size limit \cite{Cardy}. This part is successfully reproduced from
the holographic dual computation in a BTZ black hole via AdS/CFT in
\cite{RT}.

By taking the limit $\ep=1-L\to 0$, we find \be S_{A}(L=1-\ep)=
\f{1}{3}\log\ep+\f{\pi}{3\beta}+\sum_{l=1}^\infty
\f{(-1)^l}{l}\left[\f{2\pi l}{\beta}\coth\left(\f{\pi
l}{\beta}\right)-2\right]\f{1}{\sinh\left(\f{\pi
l}{\beta}\right)}.\ee Thus we can extract the finite part \be
S(1)_{finite}\equiv
S(1-\ep)-S(\ep)=\f{\pi}{3\beta}+\sum_{l=1}^\infty
\f{(-1)^l}{l}\left[\f{2\pi l}{\beta}\coth\left(\f{\pi
l}{\beta}\right)-2\right]\f{1}{\sinh\left(\f{\pi l}{\beta}\right)}.
\label{dif}\ee Clearly, the leading term $\f{\pi}{3\beta}$
represents the thermal entropy in the high temperature limit
$\beta\to 0$.

On the other hand, the full expression of thermal entropy
$S_{thermal}$ is given by \ba S_{thermal}&=&-\f{\de F}{\de
T}=\beta^2\f{\de}{\de\beta}[-\beta^{-1}\log Z]\no &=&
\f{\pi}{3\beta}+4\sum_{m=1}^\infty
\log(1+e^{-\f{2\pi}{\beta}(m-\f{1}{2})})-\f{8\pi}{\beta}\sum_{m=1}^\infty
\f{m-\f12}{e^{\f{2\pi}{\beta}(m-\f12)}+1},\label{thermals} \ea where
the partition function $Z$ is defined by \be
Z=\f{|\theta_3(0|\tau)|^2}{|\eta(\tau)|^2}=\f{|\theta_3(0|-\tau^{-1})|^2}{|\eta(-\tau^{-1})|^2}
=e^{\f{\pi}{6\beta}}\prod_{m=1}^\infty
(1+e^{-\f{2\pi}{\beta}(m-\f{1}{2})})^4.\ee Remarkably, we can show
that the total expression of (\ref{dif}) indeed agrees\footnote{
This proof is elementary.} with the thermal entropy $S_{thermal}$
for arbitrary $\beta$ \be S(1)_{finite}=S_{thermal}. \ee This
relation is very clear in the holographic picture based on AdS/CFT
as will be explained in section \ref{ss:HolInt}.

\subsection{Low Temperature Expansion}
\hspace{5mm} On the other hand, it is possible to perform the low
temperature expansion with the modular transformation undone. In the
end, we obtain similarly to (\ref{eq:EEnu=3}) \ba S_2&=&
-2\f{\de}{\de N}\sum_{k=-\f{N-1}{2}}^{\f{N-1}{2}}\sum_{m=1}^\infty
\log\left[\f{(1+e^{2\pi i\f{kL}{N}}e^{-2\pi
\beta(m-1/2)})(1+e^{-2\pi i\f{kL}{N}}e^{-2\pi
\beta(m-1/2)})}{(1+e^{-2\pi \beta(m-1/2)})^2}\right]\Biggr|_{N=1}\no
&=&2 \sum_{l=1}^\infty \f{(-1)^{l-1}}{l}\cdot \f{1-\pi lL\cot(\pi
Ll)}{\sinh(\pi l\beta)}.
 \ea

In summary, the total expression of entanglement entropy in the low
temperature expansion becomes \ba S_{A}&=&\f{1}{3}\log[\f{1}{\pi a}\sin(\pi
L)]+\f{1}{3}\sum_{m=1}^\infty \log\left[\f{(1-e^{2\pi iL}e^{-2\pi
\beta m})(1-e^{-2\pi iL}e^{-2\pi \beta m})}{(1-e^{-2\pi \beta
m})^2}\right]\no  &&\quad + 2\sum_{l=1}^\infty
\f{(-1)^{l-1}}{l}\cdot\f{1-\pi lL\cot(\pi Ll)}{\sinh(\pi l\beta)}.
\label{LOW} \ea

At zero temperature, the formula (\ref{LOW}) is simply reduced to
\be S_A=\f{c}{3}\log \left[\f{1}{\pi a}\sin(\pi L)\right], \ee and
this reproduces\footnote{Remember that we assume the space
coordinate is compactified on a circle whose length is $1$.} the
known result \cite{Cardy}. This part is successfully reproduced from
the holographic dual computation via AdS/CFT in \cite{RT}.

Still one may worry if there are many poles which come from the
final term in (\ref{LOW}). However, this turns out to be an artifact
of the order of the summation as we will see in the next subsection.
Indeed, the high and low temperature expansion will be proved to be
equivalent as they should be. The high temperature expression is
suitable for numerical computations.

\subsection{Comparison of High and Low Temperature
Expansion} \hspace{5mm} Originally, the low and high temperature
expressions of entanglement entropy come from the same two point
function (via the modular transformation) and thus they are at least
formally equivalent. However, as we have mentioned, they do not
appear to be so at first sight.

In spite of this, we can show that when they are expanded with
respect to the powers of $L$ like \ba S_{H}&=&\sum_{n=1}^\infty
C^{H}_{n}(\beta)~L^{2n}, \no
 S_{L}&=&\sum_{n=1}^\infty
C^{L}_{n}(\beta)~L^{2n},\ea each coefficient agrees with each other
i.e.\ $C^{H}_{n}(\beta)=C^{L}_{n}(\beta)$. Thus the point is the
order of summations.

\begin{figure}[htbp]
\begin{center}
\begin{tabular}{ccc}
\begin{minipage}{5.5cm}
  \begin{center}
      \includegraphics[scale=0.5]{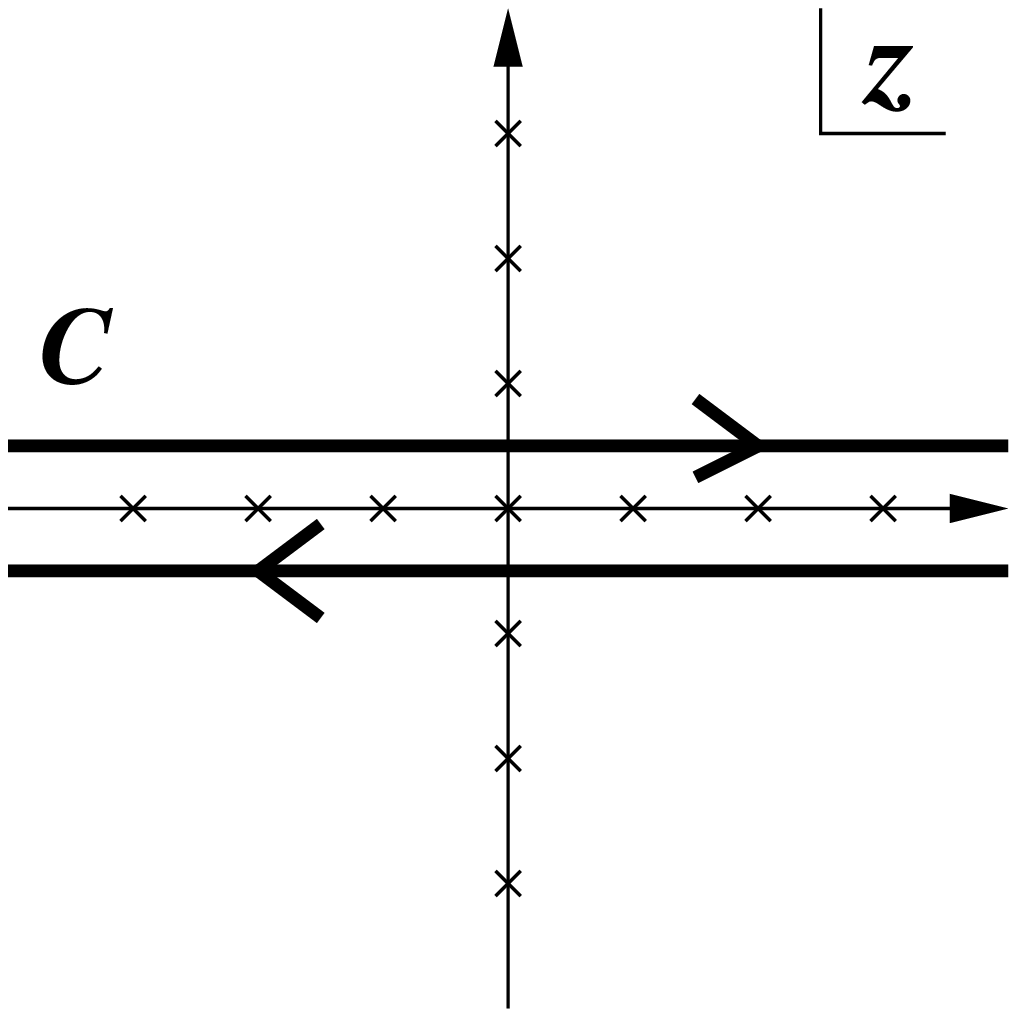}
  \end{center}
\end{minipage}
& $\xrightarrow{Contour~ Deformation}$ &
\begin{minipage}{5.5cm}
  \begin{center}
    \includegraphics[scale=0.5]{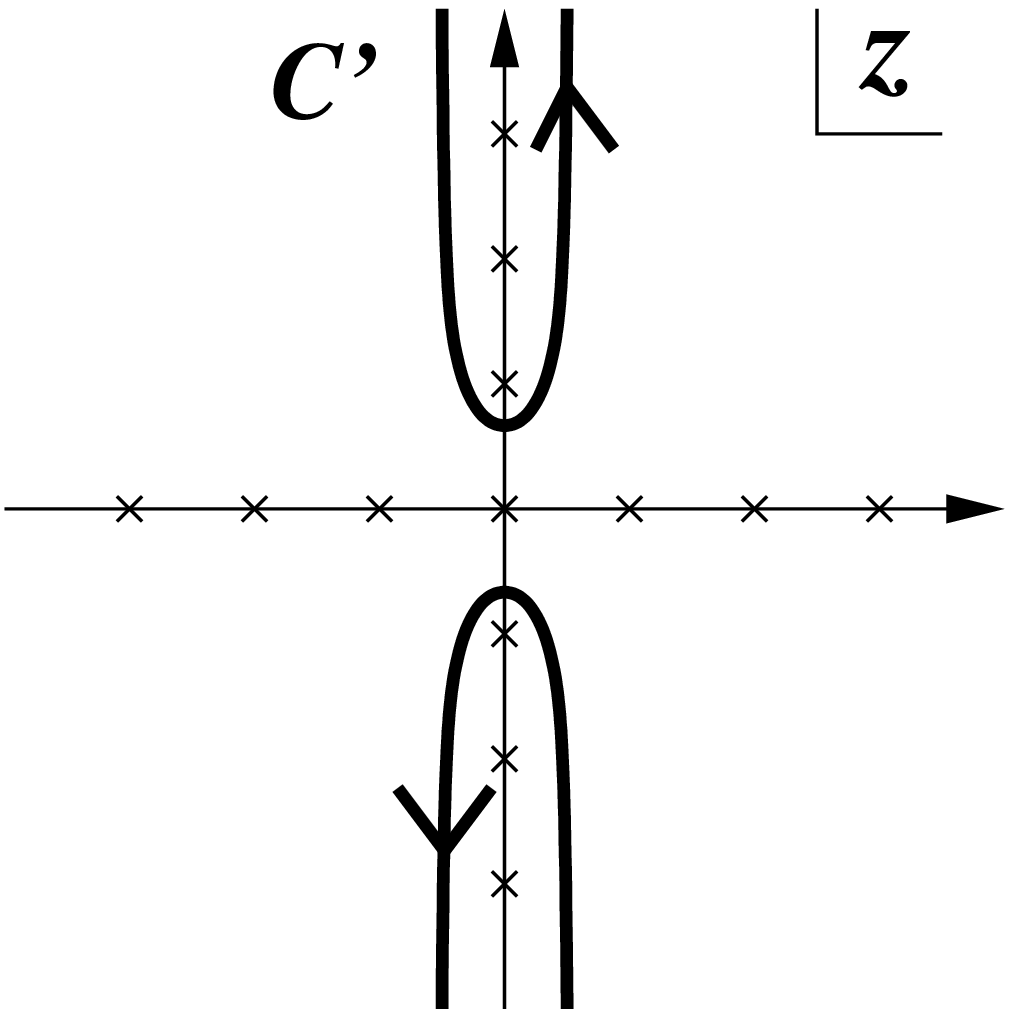}
  \end{center}
\end{minipage}
\end{tabular}
\end{center}
\caption{(\ref{eq:F_n}) can be computed in two ways; the original
  contour $C$ [Left] and the deformed contour $C'$ [Right]. The pole
  $z=0$ should be taken into account only for $n=1$ case.
} \label{fig:Contour}
\end{figure}

Let us present the proof of the equivalence. By applying the series
expansions ($B_r$ are Bernoulli numbers) \be
1-\f{x}{2}\cot\f{x}{2}=\sum_{r=1}^\infty \f{B_r}{(2r)!} x^{2r},\qquad
\f{x}{2}\coth\f{x}{2}-1=\sum_{r=1}^\infty \f{B_r(-1)^{r-1}}{(2r)!}
x^{2r}, \ee to (\ref{LOW}) and (\ref{toth}), the equalities
$C^{H}_{n}(\beta)=C^{L}_{n}(\beta)$ are rewritten as follows \ba
\f{\pi}{3\beta}+\f{2\pi^2}{3}\sum_{l=1}^\infty\f{(-1)^l\cdot
l}{\beta^2\sinh\f{\pi
l}{\beta}}&=&\f{2\pi^2}{3}\sum_{l=1}^\infty\f{(-1)^{l-1}\cdot
l}{\sinh (\pi l \beta)},\no  \sum_{l=1}^\infty \f{(-1)^{l+n-1}\cdot
l^{2n-1}}{\beta^{2n}\sinh\f{\pi l}{\beta}}&=&\sum_{l=1}^\infty
\f{(-1)^{l-1}\cdot l^{2n-1}}{\sinh(\pi l \beta)}\qquad (n\geq 2).
\ea These are equivalent to the relations \be
F_1(x)=-F_1\left(\f{1}{x}\right)+\f{1}{2\pi},\qquad F_{n}(x)=(-1)^n
\cdot F_{n}\left(\f{1}{x}\right) \ \ \ \ (n\geq 2), \label{formm}\ee
where we defined \be F_{n}(x)=\sum_{l=1}^\infty (-1)^{l-1}\cdot
l^{2n-1}\cdot \f{x^n}{\sinh(\pi lx)}. \label{form}\ee
They can be proven by considering the integral representation
\be\label{eq:F_n} F_{n}(x)=\f{1}{2\pi i}\oint_{C} dz \f{-\pi
x^{n}z^{2n-1}}{\sinh(\pi xz)\sin(\pi z)},\ee where $C$ represents
the path $z\in [-\infty+i\ep,\infty+i\ep]\cup
[\infty-i\ep,-\infty-i\ep]$ (Figure \ref{fig:Contour}). It is easy
to show (\ref{form}) by summing over the residues of poles $z\in
{\bf Z}$.

 By deforming $C$ into $C'$ which surrounds the poles
on the imaginary axis $z\in \f{i}{x}{\bf Z}\neq 0$ , we can indeed
prove (\ref{formm}) directly (only when $n=1$ we need to take into
account the pole at $z=0$).

In this way we have found that the low and high temperature
expansion are equivalent. For the actual computation the high
temperature expansion is more useful.

\subsection{Generalization}
\hspace{5mm} Here we would like to generalize the above result to
the case in which the interval $A$ extends not only in the spacial
direction but also in the temporal direction by setting $z=L+iT$,
where $T$ is the Euclidean time. We also treat $\tau=\alpha+i\beta$
as a general complex number so that it includes the rotating black
holes after the Lorentzian continuation. Remarkably, the
entanglement entropy becomes the sum of the holomorphic contribution
and the anti-holomorphic one as the thermal entropy does.

Generalization is straightforward since we have only to replace
$i\beta\to\tau=\alpha+i\beta$ and $L\to z=L+iT$ in the previous
results. The two point function of twist operators becomes \be \la
\sigma_k(z,\bar{z})\sigma_{-k}(0,0)\lb
=\left|\f{2\pi\eta(\tau)^3}{\theta_1(z|\tau)}\right|^{4\Delta_k}\cdot
\f{\theta_\nu\left(\f{k}{N}z
|\tau\right)\overline{\theta_\nu\left(\f{k}{N}z|\tau\right)}}
{|\theta_\nu(0|\tau)|^2}.\ee
In the following calculation, we restrict to $\nu=3$ as above.

We firstly evaluate $S$ in the high
temperature expansion. $S_1$ is \ba
S_{1}&=&\frac{1}{6}\log\left[-\frac{i\tau}{2\pi} e^{-\frac{\pi
iz^2}{\tau}}
\frac{\theta_{1}(\frac{z}{\tau}|-\frac{1}{\tau})}{\eta(-\frac{1}{\tau})^3}\right]
+(c.c.)\no &=& \frac{1}{6}\left [\frac{\pi
z^2}{i\tau}+\log\left[-\frac{i\tau}{\pi} \sin\left(\frac{\pi
z}{\tau} \right)\right ]+\left(\cdots\right) \right]+(c.c.), \ea
where $\cdots$ represents \be (\cdots) =\displaystyle{\sum_{m=1}
^{\infty}}\log\left[\frac{(1-e^{\frac{2\pi iz}{\tau}} \tilde{q}^m)
(1-e^{\frac{-2\pi iz}{\tau}}\tilde{q}^m)}{(1-\tilde{q}^m)^2}\right],
\ee with $\tilde{q}=e^{-\frac{2\pi i}{\tau}}$ and $(c.c.)$ is
complex conjugate of the first term which comes from the
anti-holomorphic part. $S_2$ is calculated as \ba S_{2}
&=&-\left[\frac{\pi}{6 i}\frac{z^2}{\tau} +\displaystyle{\sum_{l=1}
^{\infty}}\frac{(-1)^{l-1}}{l}\frac{ \frac{i\pi
lz}{\tau}\coth(\frac{i\pi lz}{\tau})-1} {\sinh(\frac{i\pi
l}{\tau})}\right]+(c.c.). \ea As a result we have \ba S_A &=&
\frac{c}{6}\Biggl[\log\biggl[\frac{\tau}{\pi a i}\sin\left(\frac{\pi
z}{\tau} \right)\biggr ]+\displaystyle{\sum_{m=1}^{\infty}}\log
\biggl[\frac{(1-e^{\frac{2\pi iz}{\tau}}\tilde{q}^m)
(1-e^{-\frac{2\pi iz}{\tau}}\tilde{q}^m)}{(1-\tilde{q}^m)}\biggr]\Biggr]
\no&&\quad -\displaystyle{\sum_{l=1}^{\infty}} \frac{(-1)^{l-1}}{l}\frac{
\frac{i\pi lz}{\tau}\coth(\frac{i\pi lz}{\tau})-1} {\sinh(\frac{i\pi
l}{\tau})} +(c.c.)\label{genhigh},\ea where we made the cut
off $a$ explicit.

The expression of $S_A$ in the low temperature expansion is also
given as \ba S_A &=&\frac{c}{6}\log\left[\frac{1}{\pi a}\sin(\pi z)
\displaystyle{\prod_{m=1} ^{\infty}\frac{(1-e^{2\pi
iz}q^m)(1-e^{-2\pi iz}q^m)}{(1-q^m)^2}} \right] \no && \quad +
\displaystyle{\sum_{l=1} ^{\infty}} \frac{(-1)^{l}}{l}\frac{1-\pi
lz\cot(\pi lz)}{\sinh(i\pi l\tau)}+(c.c.), \ea where $q=e^{2\pi
i\tau}$. Here the first term and the second one are contributions
from the holomorphic part of $S_1$ and $S_2$ respectively.

\subsection{Other Spin Structures}\label{otherspinstructure}
\hspace{5mm} It is also useful to find the entanglement entropy for
other spin structures of the Dirac fermions. First consider the case
of $\nu=2$ i.e.\ the finite temperature theory with the periodic
boundary condition (R sector). To calculate the entanglement entropy
in the high temperature expansion, we again apply the modular
transformation and obtain (the other parts are the same as $\n=3$
case) \be \tilde S_2 = 2 \sum_{l=1}^\infty\f{1}{l}\f{\f{\pi
     Ll}{\beta}\coth\f{\pi Ll}{\beta}-1}{\sinh\f{\pi Ll}{\beta}}.
\ee In this case, the thermal entropy defined
by (\ref{thermals}) becomes \be S_{thermal} =
\f{\pi}{3\beta} + 4\sum_{m=1}^\infty \log
(1-e^{-\f{2\pi}{\beta}(m-\f{1}{2})}) -
\f{8\pi}{\beta}\sum_{m=1}^\infty
\f{m-\half}{e^{\f{2\pi}{\beta}(m-\f{1}{2})}-1}, \ee and we can check
$S_{finite}(L=1)$ agrees with this.

It is also possible to compute the entanglement entropy in the
$\nu=4$ case. This corresponds to the index calculation in the NS
sector Tr$_{NS}(-1)^F$ and is not related any realistic thermal
distribution. In this case, similarly we obtain \be\label{eq:EEnu=4}
\tilde S_2 = -\f{\pi L}{\beta} + 2\log 2 + 4\pi \sum_{l=1}^\infty
\f{(-)^l\cdot L}{\beta(e^{\f{2\pi lL}{\beta}}-1)}+4
\sum_{l=1}^\infty \f{(-)^l}{l}\cdot\f{\f{\pi Ll}{\beta}\coth\f{\pi
Ll}{\beta}-1}{e^{\f{2\pi l}{\beta}}-1}. \ee In $\beta \to \infty$
limit, (\ref{eq:EEnu=3}) and (\ref{eq:EEnu=4}) vanish respectively.
This implies the boundary condition in the thermal direction can be
neglected in this limit as expected. The thermal entropy is \be S_{thermal} = 2\log
2-\f{2\pi}{3\beta}+4\sum_{m=1}^\infty \log (1+e^{-\f{2\pi
m}{\beta}})-\f{8\pi}{\beta}\sum_{m=1}^\infty \f{m}{e^{\f{2\pi
m}{\beta}}-1}, \ee and we can check that $S_{finite}(L=1)$ agrees
with this.

\subsection{Temporal Entanglement Entropy: Beyond the Horizon}

\begin{figure}[htbp]
\begin{center}
  \includegraphics[height=5cm]{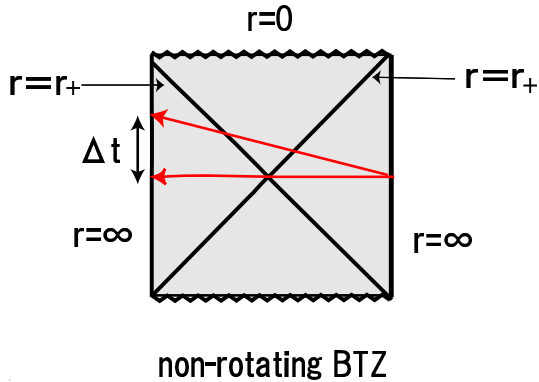}\\
  \caption{Penrose diagram of the non-rotating BTZ black hole.
  Red lines are the geodesics between two boundaries.
  }\label{norot}
\end{center}
\end{figure}

\hspace{5mm} For simplicity, here, we take $\tau = i\beta$ which
corresponds to the case of non-rotating black hole. It is worth
while to take some notice the case in which $z=\Delta
t-i\frac{\beta}{2}$ in the above generalization. The imaginary shift
$t\to t-i\frac{\beta}{2}$ of the Lorentzian time takes us from a
boundary to the other boundary \cite{Kraus, KOS} (see Figure
\ref{norot}).

When $\beta$ is sufficiently small, using the high temperature
expansion, we find \ba S_{A} \simeq
\frac{c}{3}\log\left[\frac{\beta}{\pi a} \cosh\left(\frac{\pi\Delta
t}{\beta}\right)\right]. \ea This entanglement entropy can be
calculated also from the bulk geodesic point of view since it is
related to the bulk geodesic distance $|\gamma|$ between the points
in which the twist operators are inserted \cite{RT} as in
(\ref{ARE}). The bulk geometry is the non-rotating BTZ black hole
and the Penrose diagram is in Figure \ref{norot}. The metric follows
from (\ref{metnon}) by taking $r_{-}=0$ and $\beta=\frac{2\pi
R^2}{r_{+}}$. The geodesic which corresponds to the above
calculation can be seen in the Figure \ref{norot}. Here we set $t=0$
at the initial point. The geodesic distance can be exactly found
\cite{KOS} \ba |\gamma| = 2R\log\left[\frac{\beta}{\pi a}\cosh\left(
\frac{\pi\Delta t}{\beta}\right)\right]. \ea Since the central
charge is given by $c=\frac{3R}{2G_{N}^{(3)}}$ \cite{BrHe}, we can
precisely show the equality $S_{ent}=S_{A}$.

We see as above that the bulk and the boundary calculations are
identical. Notice that the geodesic involved in the bulk computation
now extends beyond the event horizon. Even though we have a definite
definition of this temporal entanglement entropy in the Euclidean
CFT, the physical meaning of this temporal entanglement entropy is
not clear. It may be an analogue of Polyakov loop in the context of
Wilson loops. Further understandings of this quantity will deserve a
future study.

\section{Conclusion and Discussion}
\hspace{5mm} In this paper, we have explored the origin of black
hole entropy
 from the viewpoint of AdS/CFT correspondence. We have been
 particularly interested in the black holes whose near horizon
 geometries include AdS$_2$. Extremal or near extremal black holes
 in 4D and 5D are falling into this class.
 We argued that the AdS$_2/$CFT$_1$ correspondence leads to the
 equivalence between the black hole entropy and the
 von-Neumann entropy associated with the quantum entanglement
 between a pair of quantum mechanical systems. The remarkable
 fact that the AdS$_2$ space in the global coordinate has two time-like
 boundaries plays a crucial role in this quantum entanglement.
 This turns out to be the reason why we get non-zero entropy of
 extremal black holes though its dual AdS$_2$ space is at zero
 temperature. This may be comparable to the entanglement
 interpretation for AdS black holes in higher dimension considered
 in \cite{ME}.

In summary, the mechanism of producing non-zero entanglement entropy
is as follows. First, the BPS states in the internal spaces (such as
Calabi-Yau spaces, $K3$ or $T^4$) produce a large degeneracy of
ground states. Then the AdS$_2$ space, which has two boundaries,
maximally entangles them and in the end we obtain a large
entanglement entropy which agrees with the black hole entropy.

There is a possibility that we have to restrict the physical
spacetime to a certain region (e.g. outside the inner horizon
\cite{ME,LR,BaLe}) in the global AdS. However, our derivation of
entanglement entropy can still be applied without any change even in
such a case, as long as there are two time-like boundaries. As we
mentioned in the last of section 3, this may lead to a subtle issue
in the strictly extremal black holes. Though we believe this is not
a serious problem, the better understanding of this subtlety as well
as the precise derivation of the two point functions (\ref{onet})
and (\ref{onest}) from the CQM side, will be important future
problems.

In the latter part of this paper, we computed the entanglement
entropy in the 2D free Dirac fermion theory. We obtained an
analytical expression in the presence of both the finite size and
finite temperature effect. This is the first analytical result of
entanglement entropy in 2D CFT which takes both effects into
account. Importantly, the result depends not only on the central
charge of the CFT but also on many other details of the theory. This
analysis enables us to show explicitly that the entanglement entropy
is reduced to the thermal entropy when the subsystem $A$ becomes
coincident with the total system. As we pointed out in section 2,
this relation offers a further evidence for the holographic
computation of entanglement entropy found in \cite{RT}), which also
plays an important role in our discussions of AdS$_2$/CFT$_1$. It is
also interesting to extend our results to a 2D free massless scalar
field theory and eventually to the symmetric orbifold theories which
have clear holographic duals.\\

\centerline{\bf Acknowledgments}

We are extremely grateful to P. Calabrese, J. Cardy, V. Hubeny, M.
Rangamani, A. Strominger and Y. Tachikawa for very helpful comments
on the draft of this paper and important suggestions. TT would like
to thank the high energy theory group in Harvard university for
hospitality where this work was finalized. The work of TT is
supported in part by JSPS Grant-in-Aid for Scientific Research
No.18840027 and by JSPS Grant-in-Aid for Creative Scientific
Research No. 19GS0219. The work of TN is
supported by JSPS Grant-in-Aid for Scientific Research
No. 19 $\cdot$ 3589.


\end{document}